# The Efficient Server Audit Problem, Deduplicated Re-execution, and the Web (extended version)*

Cheng Tan, Lingfan Yu, Joshua B. Leners*, and Michael Walfish

NYU Department of Computer Science, Courant Institute       *Two Sigma Investments

**Abstract**

You put a program on a concurrent server, but you don't trust the server; later, you get a trace of the actual requests that the server received from its clients and the responses that it delivered. You separately get logs from the server; these are untrusted. How can you use the logs to efficiently *verify* that the responses were derived from running the program on the requests? This is the *Efficient Server Audit Problem*, which abstracts real-world scenarios, including running a web application on an untrusted provider. We give a solution based on several new techniques, including simultaneous replay and efficient verification of concurrent executions. We implement the solution for PHP web applications. For several applications, our verifier achieves 5.6–10.9× speedup versus simply re-executing, with <10% overhead for the server.

## 1 Motivation and contents

Dana the Deployer works for a company whose employees use an open-source web application built from PHP and a SQL database. The application is critical: it is a project management tool (such as JIRA), a wiki, or a forum. For convenience and performance, Dana wants to run the application on a cloud platform, say AWS [1]. However, Dana has no visibility into AWS. Meanwhile, undetected corrupt execution—as could happen from misconfiguration, errors, compromise, or adversarial control at any layer of the execution stack: the language run-time, the HTTP server, the OS, the hypervisor, the hardware—would be catastrophic for Dana's company. So Dana would like assurance that AWS is executing the actual application as written. How can Dana gain this assurance?

Dana's situation is one example of a fundamental problem, which this paper defines and studies: the *Efficient Server Audit Problem*. The general shape of this problem is as follows. A principal supplies a *program* to an untrusted *executor* that is supposed to perform repeated and possibly concurrent executions of the program, on different inputs. The principal later wants to *verify* that the outputs delivered by the executor were produced by running the program. The verification algorithm, or *verifier*, is given an accurate *trace* of the executor's inputs and delivered outputs. In addition, the executor gives the verifier *reports*, but these are untrusted and possibly spurious. The verifier must somehow use the reports to determine whether the outputs in the trace are consistent with having actually executed the program. Furthermore, the verifier must make this determination efficiently; it should take less work than re-executing the program on every input in the trace.

The requirement of a trace is fundamental: if we are auditing a server's outputs, then we need to know those outputs. Of course, getting a trace may not be feasible in all cases. In Dana's case, the company can place a middlebox at the network border, to capture end-clients' traffic to and from the application. We discuss other scenarios later (§4.1, §7).

We emphasize that the Efficient Server Audit Problem is separate from—but complementary to—program verification, which is concerned with developing bug-free programs. Our concern instead is whether a given program is actually executed as written. Neither guarantee subsumes the other.

The high-level consideration here is execution integrity, a topic that has been well-studied in several academic communities, with diverse solutions (§6.1). The novelty in our variant is in combining three characteristics: (1) we make no assumptions about the failure modes of the executor, (2) we allow the executor to be concurrent, and (3) we insist on solutions that scale beyond toy programs and are compatible with (at least some) legacy programs.

The contributions and work of this paper are as follows.

§2 **Definition of the Efficient Server Audit Problem.** We first present the problem in theoretical terms. We do this to show the generality and the fundamental challenges.

§3 **An abstract solution: SSCO.** We exhibit a solution at a theoretical level, so as to highlight the core concepts, techniques, and algorithms. These include:

§3.1 **SIMD [6]-on-demand.** The verifier re-executes all requests, in an accelerated way. For a group of requests with the same control flow, the verifier executes a "superposition": instructions with identical operands across requests are performed once, whereas instructions with different operands are executed individually and merged into the superposed execution. This solution assumes that the workload has repeated traversal of similar code paths—which is at least the case for some web applications, as observed by Poirot [53, §5].

§3.3 **Simulate-and-check.** How can the verifier re-execute *reads* of persistent or shared state? Because it re-

---

executes requests out of order, it cannot physically re-invoke operations on such state, but neither can it trust reports that are allegedly the originally read values (§3.2). Instead, the executor (purportedly) logs each operation's operands; during re-execution, the verifier simulates reads, using the writes in the logs, and checks the logged writes opportunistically.

§3.5 **Consistent ordering.** The verifier must ensure that operations can be consistently ordered (§3.4). To this end, the verifier builds a directed graph with a node for every external observation or alleged operation, and checks whether the graph is acyclic. This step incorporates an efficient algorithm for converting a trace into a time precedence graph. This algorithm would accelerate prior work [14, 51] and may be useful elsewhere.

SSCO has other aspects besides, and the unified whole was difficult to get right (§7): our prior attempts had errors that came to light when we tried to prove correctness. This version, however, is proved correct (Appendix A).

§4 **A built system: OROCHI.** We describe a system that implements SSCO for PHP web applications. This is for the purpose of illustration, as we expect the system to generalize to other web languages, and the theoretical techniques in SSCO to apply in other contexts (§7). OROCHI includes a record-replay system [34, 35] for PHP [27, 53]. The replayer is a modified language runtime that implements SIMD-on-demand execution using *multivalue* types that hold the program state for multiple re-executions. OROCHI also introduces mechanisms, based on a versioned database [27, 41, 62, 81], to adapt simulate-and-check to databases and to deduplicate database queries.

§5 **Experimental evaluation of OROCHI.** In experiments with several applications, the verifier can audit 5.6–10.9× faster than simple re-execution; this is a loose lower bound, as the baseline is very pessimistic for OROCHI (§5.1). OROCHI imposes overhead of roughly 10% on the web server. OROCHI's reports, per-request, are 3%–11% of the size of a request-response pair. Most significantly, the verifier must keep a copy of the server's persistent state.

The main limitations (§5.5) are that, first, in SSCO the executor has discretion over scheduling concurrent requests, and it gets additional discretion, in OROCHI, over the return values of non-deterministic PHP built-ins. Second, OROCHI is restricted to applications that do not interact much with other applications; nevertheless, there are suitable application classes, for example LAMP [3]. Third, OROCHI requires minor modifications in some applications, owing to the SSCO model. Finally, the principal can audit an application only after activating OROCHI; if the server was previously running, the verifier has to bootstrap from the pre-OROCHI state.

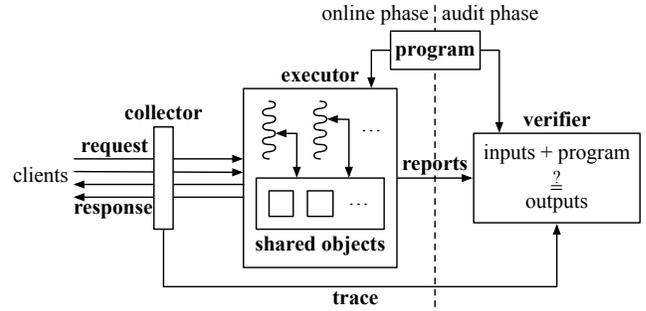

**Figure 1**—The Efficient Server Audit Problem. The objects abstract shared state (databases, key-value stores, memory, etc.). The technical problem is to design the verifier and the reports to enable the verifier, given a trace and a program, to efficiently validate (or invalidate) the contents of responses.

## 2 Problem definition

This section defines the Efficient Server Audit Problem. The actors and components are depicted in Figure 1.

A *principal* chooses or develops a *program*, and deploys that program on a powerful but untrusted *executor*.

Clients (the outside world) issue *requests* (inputs) to the executor, and receive *responses* (outputs). A response is supposed to be the output of the program, when the corresponding request is the input. But the executor is untrusted, so the response could be anything.

A *collector* captures an ordered list, or *trace*, of requests and responses. We assume that the collector does its job accurately, meaning that the trace exactly records the requests and the (possibly wrong) responses that actually flow into and out of the executor.

The executor maintains *reports* whose purpose is to assist an audit; like the responses, the reports are untrusted.

Periodically, the principal conducts an audit; we often refer to the audit procedure as a *verifier*. The verifier gets a trace (from the accurate collector) and reports (from the untrusted executor). The verifier needs to determine whether executing the program on each input in the trace truly produces the respective output in the trace.

Two features of our setting makes this determination challenging. First, the verifier is much weaker than the executor, so it cannot simply re-execute all of the requests.

The second challenge arises from concurrency: the executor is permitted to handle multiple requests at the same time (for example, by assigning each to a separate thread), and the invoked program is permitted to issue *operations* to *objects*. An object abstracts state shared among executions, for example a database, key-value store, or memory cells (if shared). We will be more precise about the concurrency model later (§3.2). For now, a key point is that, given a trace—in particular, given the ordering of requests and responses in the trace, and given the contents of requests—the number of valid possibilities for the contents of responses could be immense. This is because an executor's responses depend on the contents of shared objects; as usual in concurrent systems, those contents depend on the operation order, which depends on the executor's internal scheduling choices.

Somehow, the reports, though unreliable, will have to help the verifier efficiently tell the difference between valid and invalid traces. In detail, the problem is to design the verifier and the reports to meet these properties:

- *Completeness.* If the executor behaved during the time period of the trace, meaning that it executed the given program under the appropriate concurrency model, then the verifier must accept the given trace.

- *Soundness.* The verifier must reject if the executor misbehaved during the time period of the trace. Specifically, the verifier accepts only if there is some schedule *S*, meaning an interleaving or context-switching among (possibly concurrent) executions, such that: (a) executing the given program against the inputs in the trace, while following *S*, reproduces exactly the respective outputs in the trace, and (b) *S* is consistent with the ordering in the trace. (Appendix A states Soundness precisely.) This property means that the executor can pass the audit only by executing the program on the received requests—or by doing something externally indistinguishable from that.

- *Efficiency.* The verifier must require only a small fraction of the computational resources that would be required to re-execute each request. Additionally, the executor's overhead must be only a small fraction of its usual costs to serve requests (that is, without capturing reports). Finally, the solution has to work for applications of reasonable scale.

We acknowledge that "small fraction" and "reasonable scale" may seem out of place in a theoretical description. But these characterizations are intended to capture something essential about the class of admissible solutions. As an example, there is a rich theory that studies execution integrity (§6.1), but the solutions (besides not handling concurrency) are so far from scaling to the kinds of servers that run real applications that we must look for something qualitatively different.

## 3 A solution: SSCO

This section describes an abstract solution to the Efficient Server Audit Problem, called SSCO (a rough abbreviation of the key techniques). SSCO assumes that there is similarity among the executions, in particular that there are a relatively small number of control flow paths induced by requests (§3.1). SSCO also assumes a certain concurrency model (§3.2).

**Overview and key techniques.** In SSCO, the reports are:

- *Control flow groupings*: For each request, the executor records an opaque tag that purportedly identifies the control flow of the execution; requests that induce the same control flow are supposed to receive the same tag.

- *Operation logs*: For each shared object, the executor maintains an ordered log of all operations (across all requests).

- *Operation counts*: For each request execution, the executor records the total number of object operations that it issued.

The verifier begins the audit by checking that the trace is balanced: every response must be associated with an earlier request, and every request must have a single response or some information that explains why there is none (a network reset by a client, for example). Also, the verifier checks that every request-response pair has a unique *requestID*; a well-behaved executor ensures this by labeling responses. If these checks pass, we (and the verifier) can refer to request-response pairs by requestID, without ambiguity.

The core of verification is as follows. The verifier re-executes each control flow group in a batch; this happens via *SIMD [6]-on-demand execution* (§3.1). During this process, re-executed object operations don't happen directly—they can't, as re-execution follows a different order from the original (§3.2). Instead, the operation logs contain a record of reads and writes, and re-execution follows a discipline that we call *simulate-and-check* (§3.3): re-executed read operations are fed (or simulated) based on the most recent write entry in the logs, and the verifier checks logged write operations opportunistically. In our context, simulate-and-check makes sense only if alleged operations can be ordered consistent with observed requests and responses (§3.4); the verifier determines whether this is so using a technique that we call *consistent ordering verification* (§3.5).

At the end, the verifier compares each request's produced output to the request's output in the trace, and accepts if and only if all of them match, across all control flow groups.

The full audit logic is described in Figures 3, 5, and 6, and proved correct in Appendix A.

### 3.1 SIMD-on-demand execution

We assume here that requests do not interact with shared objects; we remove that assumption in Section 3.2. (As we have just done, we will sometimes use "request" as shorthand for "the execution of the program when that request is input.")

The idea in SIMD-on-demand execution is that, for each control flow group, the verifier conducts a single "superposed" execution that logically executes all requests in that group together, at the same time. Instructions whose operands are different across the separate logical executions are performed separately (we call this *multivalent* execution of an instruction), whereas an instruction executes only once (*univalently*) if its operands are identical across the executions. The concept is depicted in Figure 2.

The control flow groupings are structured as a map *C* from opaque tag to set-of-requestIDs. Of course, the map is part of the untrusted report, so the verifier does not trust it. However, if the map is incorrect (meaning, two requests in the same control flow group diverge under re-execution), then the verifier rejects. Furthermore, if the map is incomplete (meaning, not including particular requestIDs), then the re-generated responses will not match the outputs in the trace. The verifier can filter out duplicates, but it does not have to do so, since re-execution is idempotent (even with shared objects, below).

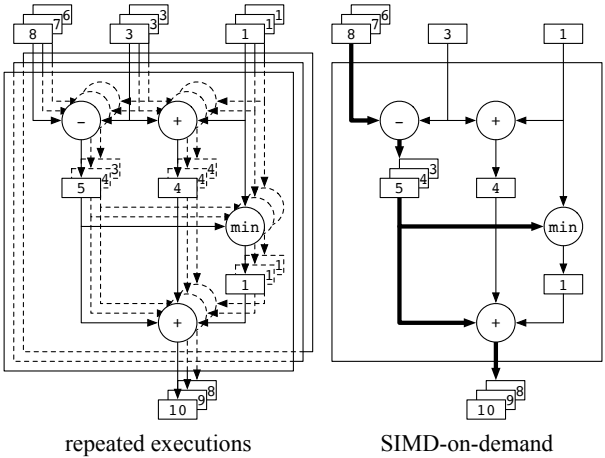

**Figure 2**—Abstract depiction of SIMD-on-demand, for a simple computation. Rectangles represent program variables, circles represent instructions. On the right, thick lines represent explicitly materialized outputs; thin lines represent collapsed outputs.

Observe that this approach meets the verifier's efficiency requirement (§2), if (1) the number of control paths taken is much smaller than the number of requests in the audit, (2) most instructions in a control flow group execute univalently, and (3) it is inexpensive to switch between multivalent and univalent execution, and to decide which to perform. (We say "if" and not "only if" because there may be platforms where, for example, condition (1) alone is sufficient.)

**System preview.** The first two conditions hold in the setting for our built system, OROCHI (§4): LAMP web applications. Condition (1) holds because these applications are in a sense routine (they do similar things for different users) and because the programming language is high-level (for example, string operations or calls like sort() or max() induce the same control flow [53]). Condition (2) holds because the logical outputs have a lot of overlap: different users wind up seeing similar-looking web pages, which implies that the computations that produce these web pages include identical data flows. This commonality was previously observed by Poirot [53], and our experiments confirm it (§5.2). Condition (3) is achieved, in OROCHI, by augmenting the language run-time with *multivalue* versions of basic datatypes, which encapsulate the different values of a given operand in the separate executions. Re-execution moves dynamically between a vector, or SIMD, mode (which operates on multivalues) and a scalar mode (which operates on normal program variables).

### 3.2 Confronting concurrency and shared objects

As noted earlier, a key question is: how does the verifier re-execute an operation that reads from a shared object? An approach taken elsewhere [27, 53] is to record the values that had been read by each request, and then to supply those values during re-execution. One might guess that, were we to apply this approach to our context where reports are untrusted, the worst thing that could happen is that the verifier would fail to reproduce the observed outputs in the trace—in other words, the executor would be incriminating itself. But the problem is much worse than that: the reported values and the responses could both be bogus. As a result, if the verifier's re-execution dutifully incorporated the purported read values, it could end up reproducing, and thereby validating, a spurious response from a misbehaved executor; this violates Soundness (§2).

**Presentation plan.** Below, we define the concurrency model and object semantics, as necessary context. We then cover the core object-handling mechanisms (§3.3–§3.5). However, that description will be incomplete, in two ways. First, we will not cover every check or justify each line of the algorithms. Second, although we will show with reference to examples why certain alternatives fail, that will be intuition and motivation, only; correctness, meaning Completeness and Soundness (§2), is actually established end-to-end, with a chain of logic that does not enumerate or reason about all the ways in which reports and responses could be invalid (Appendix A).

**Concurrency model and object semantics.** In a well-behaved executor, each request induces the creation of a separate *thread* that is destroyed after the corresponding response is delivered. A thread runs concurrently with the threads of any other requests whose responses have not yet been delivered. Each thread sequentially performs instructions against an isolated execution context: registers and local memory. As stated earlier, threads perform *operations* on shared objects (§2). These operations are blocking, and the objects expose atomic semantics. We assume for simplicity in this section that objects expose a read-write interface; they are thus atomic registers [55]. Later, we will permit more complex interfaces, such as SQL transactions (§4.4).

### 3.3 Simulate-and-check

The reports in SSCO include the (alleged) operations themselves, in terms of their operands. Below, we describe the format and how the verifier uses these operation logs.

**Operation log contents.** Each shared object is labeled with an index $i$. The operation log for object $i$ is denoted $OL_i$, and it has the following form ($\mathbb{N}^+$ denotes the set $\{1, 2, \ldots\}$):

$$OL_i \colon \mathbb{N}^+ \to (\text{requestID}, \text{opnum}, \text{optype}, \text{opcontents}).$$

The opnum is per-requestID; a correct executor tracks and increments it as requestID executes. An operation is thus

```
Input Trace Tr    Input Reports R    Global OpMap: (requestID, opnum) → (i, seqnum)
    Components of the reports R:                                           24: procedure REEXEC()
C: CtlFlowTag → Set(requestIDs) // purported groups; §3.1                  25:    Re-execute Tr in groups according to C:
OL_i: ℕ⁺ → (requestID, opnum, optype, opcontents) // purported op logs; §3.3   26:
M: requestID → ℕ    // purported op counts; §3.3                           27:        (1) Initialize a group as follows:
                                                                           28:            Read in inputs for all requests in the group
                                                                           29:            Allocate program structures for each request in the group
 1: procedure SSCO_AUDIT()                                                 30:            opnum ← 1    // opnum is a per-group running counter
 2:     // Partially validate reports (§3.5) and construct OpMap            31:
 3:     ProcessOpReports()    // defined in Figure 5                       32:        (2) During SIMD-on-demand execution (§3.1):
 4:                                                                        33:
 5:     return ReExec()    // line 24                                      34:            if execution within the group diverges: return REJECT
 6:                                                                        35:
 7: procedure CHECKOP(rid, opnum, i, optype, opcontents)                   36:            When the group makes a state operation:
 8:     if (rid, opnum) not in OpMap: REJECT                               37:                optype ← the type of state operation
 9:                                                                        38:                for all rid in the group:
10:     î, s ← OpMap[rid, opnum]                                           39:                    i, oc ← state op parameters from execution
11:     ôt, ôc ← (OL_i[s].optype, OL_i[s].opcontents)                      40:                    s ← CheckOp(rid, opnum, i, optype, oc) // line 7
12:     if i ≠ î or optype ≠ ôt or opcontents ≠ ôc:                        41:                    if optype = RegisterRead:
13:         REJECT                                                         42:                        state op result ← SimOp(i, s, optype, oc) // line 16
14:     return s                                                           43:                opnum ← opnum + 1
15:                                                                        44:
16: procedure SIMOP(i, s, optype, opcontents)                              45:        (3) When a request rid finishes:
17:     ret ← ⊥                                                            46:            if opnum < M(rid): return REJECT
18:     writeop ← walk backward in OL_i from s; stop when                  47:
19:         optype=RegisterWrite                                           48:        (4) Write out the produced outputs
20:     if writeop doesn't exist:                                          49:
21:         REJECT                                                         50:    if the produced outputs from (4) are exactly the responses in Tr:
22:     ret = writeop.opcontents                                           51:        return ACCEPT
23:     return ret                                                         52:    return REJECT
```

**Figure 3**—The SSCO audit procedure. The supplied trace $Tr$ must be balanced (§3), which the verifier ensures before invoking SSCO_AUDIT. A rigorous proof of correctness is in Appendix A.

identified with a unique ($rid$, $opnum$) pair. The optype and opcontents depend on the object type. For registers, optype can be RegisterRead (and opcontents are supposed to be empty) or RegisterWrite (and opcontents is the value to write).

**What the verifier does.** The core re-execution logic is contained in ReExec (Figure 3, line 24). The verifier feeds re-executed *reads* by identifying the latest write before that read in the log. Of course, the logs might be spurious, so for *write* operations, the verifier opportunistically checks that the operands (produced by re-execution) match the log entries.

In more detail, when re-executing an operation ($rid, opnum$), the verifier uses *OpMap* (as defined in Fig. 3) to identify the log entry; it then checks that the parameters (generated by program logic) match the logs. Specifically, the verifier checks that the targeted object corresponds to the (unique) log that holds ($rid, opnum$) (uniqueness is ensured by checks in Figure 5), and that the produced operands (such as the value to be written) are the same as in the given log entry (lines 37–40, Figure 3). If the re-executed operation is a read, the verifier feeds it by identifying the write that precedes ($rid, opnum$); this is done in SimOp.

Notice that an operation that reads a given write might re-execute long before the write is validated. The intuition here is that a read's validity is contingent on the validity of all prior write operations in the log. Meanwhile, the audit procedure succeeds only if all checks—including the ones of write operations—succeed, thereby retroactively discharging the assumption underlying every read.

What prevents the executor from justifying a spurious response by inserting into the logs additional operations? Various checks in the algorithm would detect this and other cases. For example, the op count reports $M$ enforce certain invariants, and interlocking checks in the algorithms validate $M$.

### 3.4 Simulate-and-check is not enough

To show why simulate-and-check is insufficient by itself, and to illustrate the challenge of augmenting it, this section walks through several simple examples. This will give intuition for the techniques in the next section (§3.5).

The examples are depicted in Figure 4 and denoted **a**, **b**, **c**. Each of them involves two requests, $r_1$ and $r_2$. Each example consists of a particular trace—or, equivalently, a particular request-response pattern—and particular reports. As a shorthand, we notate the delivered responses with a pair (r1resp, r2resp); for example, the responses in **a** are $(1, 0)$.

A correct verifier must reject **a**, reject **b**, and accept **c**.

To see why, note that in **a**, the executor delivers a response to $r_1$ before $r_2$ arrives. So the executor must have executed $r_1$

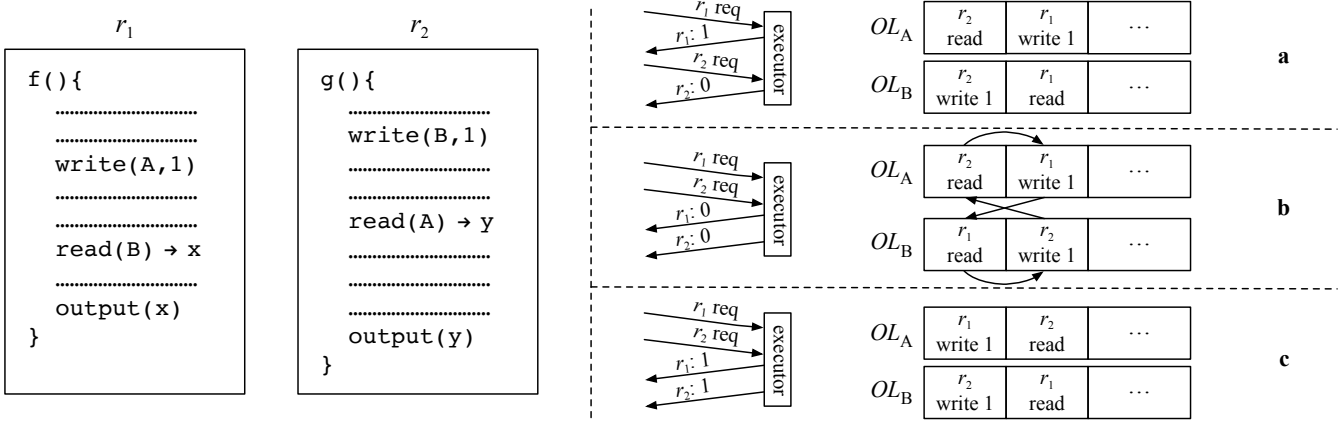

**Figure 4**—Three examples to highlight the verifier's challenge and to motivate consistent ordering verification (§3.5). As explained in the text, a correct verifier (meaning Complete and Sound; §2) must reject examples **a** and **b**, and accept **c**. In these examples, $r_1$ and $r_2$ are requestIDs in different control flow groups, and their executions invoke different subroutines of the given program. For simplicity, there is only one request per control flow group, and objects are assumed to be initialized to 0. What varies among examples are the timing of requests and responses, the contents of the executor's responses, and the alleged operation logs for objects $A$ and $B$ (denoted $OL_A$, $OL_B$). The opnum component of the log entries is not depicted.

and then executed $r_2$. Under that schedule, there is no way to produce the observed output $(1, 0)$; in fact, the only output consistent with the observed events is $(0, 1)$. Thus, accepting **a** would violate Soundness (§2).

In **b**, $r_1$ and $r_2$ are concurrent. A well-behaved executor can deliver any of $(0, 1)$, $(1, 0)$, or $(1, 1)$, depending on the schedule that it chooses. Yet, the executor delivered $(0, 0)$, which is consistent with no schedule. So accepting **b** would also violate Soundness.

In **c**, $r_1$ and $r_2$ are again concurrent. This time, the executor delivered $(1, 1)$, which a well-behaved executor can produce, by executing the two writes before either read. Therefore, rejecting **c** would violate Completeness (§2).

Now, if the verifier used only simulate-and-check (Figure 3), the verifier would accept in all three of the examples. We encourage curious readers to convince themselves of this behavior by inspecting the verifier's logic and the examples. Something to note is that in **a** and **b**, the operation logs and responses are both spurious, but they are arranged to be consistent with each other.

Below are some strawman attempts to augment simulate-and-check, by analyzing all operation logs prior to re-execution.

- What if the verifier (i) creates a global order $O$ of requests that is consistent with the real-time order (in **a**, $r_1$ would be prior to $r_2$ in $O$; in **b** and **c**, either order is acceptable), and (ii) *for each log*, checks that the order of its operations is consistent with $O$? This would rightly reject **a** ($r_1$ is before $r_2$ in $O$ but not in the logs), rightly reject **b** (regardless of the choice of $O$, one of the two logs will violate it), and wrongly reject **c** (for the same reason it would reject **b**). This approach would be tantamount to insisting that entire requests execute atomically (or transactionally)—which is contrary to the concurrency model.

- What if the verifier creates only a partial order $O'$ on requests that is consistent with the real-time order, and then insists that, for each log, the order of operations is consistent with $O'$? That is, operations from concurrent requests can interleave in the logs. This would rightly reject **a** and rightly accept **c**. But it would wrongly accept **b**.

- Now notice that the operations in **b** cannot be ordered: considering log and program order, the operations form a cycle, depicted in Figure 4. So what if the verifier (a) creates a directed graph whose nodes are all operations in the log and whose edges are given by log order and program order, and (b) checks that there are no cycles? That would rightly reject **b** and accept **c**. But it would wrongly accept **a**.

The verifier's remaining techniques, described next, can be understood as combining the preceding failed attempts.

### 3.5 Consistent ordering verification

At a high level, the verifier *ensures the existence of an implied schedule that is consistent with external observations and alleged operations*. Prior to re-executing, the verifier builds a directed graph $G$ with a node for every *event* (an observed request or response, or an alleged operation); edges represent precedence [55]. The verifier checks whether $G$ is acyclic. If so, then all events can be consistently ordered, and the implied schedule is exactly the ordering implied by $G$'s edges. Note, however, that the verifier does not follow that order when re-executing nor does the verifier consult $G$ again.

Figures 5 and 6 depict the algorithms. $G$ contains nodes labeled (*rid*, *opnum*), one for each alleged operation in the logs. $G$ also contains, for each request *rid* in the trace, nodes

```
 1: Global Trace Tr, Reports R, Graph G, OpMap OpMap
 2: procedure PROCESSOPREPORTS()
 3:
 4:     G_Tr ← CreateTimePrecedenceGraph()    // defined in Figure 6
 5:     SplitNodes(G_Tr)
 6:     AddProgramEdges()
 7:
 8:     CheckLogs()          // also builds the OpMap
 9:     AddStateEdges()
10:
11:     if CycleDetect(G):   // standard algorithm; see [32, Ch. 22]
12:         REJECT
13:
14: procedure SPLITNODES(Graph G_Tr)
15:     G.Nodes ← {}, G.Edges ← {}
16:     for each node rid ∈ G_Tr.Nodes:
17:         G.Nodes += { (rid, 0), (rid, ∞) }
18:     for each edge ⟨rid_1, rid_2⟩ ∈ G_Tr.Edges:
19:         G.Edges += ⟨(rid_1, ∞), (rid_2, 0)⟩
20:
21: procedure ADDPROGRAMEDGES()
22:     for all rid that appear in the events in Tr:
23:         for opnum = 1, …, R.M(rid):
24:             G.Nodes += (rid, opnum)
25:             G.Edges += ⟨(rid, opnum − 1), (rid, opnum)⟩
26:         G.Edges += ⟨(rid, R.M(rid)), (rid, ∞)⟩
27:
28: procedure CHECKLOGS()
29:     for log = R.OL_1, …, R.OL_n:
30:         for j = 1, …, length(log):
31:             if log[j].rid does not appear in Tr or
32:                 log[j].opnum ≤ 0 or
33:                 log[j].opnum > R.M(log[j].rid) or
34:                 (log[j].rid, log[j].opnum) is in OpMap:
35:                 REJECT
36:
37:             let curr_op = (log[j].rid, log[j].opnum)
38:             OpMap[curr_op] ← (i, j) // i is the index such that log = R.OL_i
39:
40:     for all rid that appear in the events in Tr:
41:         for opnum = 1, …, R.M(rid):
42:             if (rid, opnum) is not in OpMap: REJECT
43:
44: procedure ADDSTATEEDGES()
45:     // Add edge to G if adjacent log entries are from different
46:     // requests. If they are from the same request, check that the
47:     // intra-request opnum increases
48:     for log = R.OL_1, …, R.OL_n:
49:         for j = 2, …, length(log):
50:             let curr_r, curr_op, prev_r, prev_op =
51:                 (log[j].rid, log[j].opnum, log[j−1].rid, log[j−1].opnum)
52:             if prev_r ≠ curr_r:
53:                 G.Edges += ⟨(prev_r, prev_op), (curr_r, curr_op)⟩
54:             else if prev_op > curr_op: REJECT
```

**Figure 5**—ProcessOpReports ensures that events (request arrival, departure of response, and operations) can be consistently ordered. It does this by constructing a graph $G$—the nodes are events; the edges reflect request precedence in $Tr$, program order, and the operation logs—and ensuring that $G$ has no cycles. $OpMap$ is constructed here as an index of the operation logs.

```
 1: procedure CREATETIMEPRECEDENCEGRAPH()
 2:     // "Latest" requests; "parent(s)" of any new request
 3:     Frontier ← {}
 4:     G_Tr.Nodes ← {}, G_Tr.Edges ← {}
 5:
 6:     for each input and output event in Tr, in time order:
 7:         if the event is REQUEST(rid):
 8:             G_Tr.Nodes += rid
 9:             for each r in Frontier:
10:                 G_Tr.Edges += ⟨r, rid⟩
11:         if the event is RESPONSE(rid):
12:             // rid enters Frontier, evicting its parents
13:             Frontier −= { r | ⟨r, rid⟩ ∈ G_Tr.Edges }
14:             Frontier += rid
15:     return G_Tr
```

**Figure 6**—Algorithm for explicitly materializing the time-precedence partial order, $<_{Tr}$, in a graph. The algorithm constructs $G_{Tr}$ so that $r_1 <_{Tr} r_2 \iff G_{Tr}$ has a directed path from $r_1$ to $r_2$. $Tr$ is assumed to be a (balanced; §3) list of REQUEST and RESPONSE events in time order.

$(rid, 0)$ and $(rid, \infty)$, representing the arrival of the request and the departure of the response, respectively. The edges in $G$ capture program order via AddProgramEdges and alleged operation order via AddStateEdges.

**Capturing time precedence.** To be consistent with external observations, $G$ must also capture time precedence. (This is what was missing in the final attempt in §3.4.) We say that $r_1$ precedes $r_2$ (notated $r_1 <_{Tr} r_2$) if the trace $Tr$ shows that $r_1$ departed from the system before $r_2$ arrived [55]. If $r_1 <_{Tr} r_2$, then the operations issued by $r_1$ must occur in the implied schedule prior to those issued by $r_2$.

Therefore, the verifier needs to construct edges that capture the $<_{Tr}$ partial order, in the sense that $r_1 <_{Tr} r_2 \iff G$ has a directed path from $(r_1, \infty)$ to $(r_2, 0)$. How can the verifier construct these edges from the trace? Prior work [14] gives an *offline* algorithm for this problem that runs in time $O(X \cdot \log X + Z)$, where $X$ is the number of requests, and $Z$ is the minimum number of time-precedence edges needed (perhaps counter-intuitively, more concurrency leads to higher $Z$).

By contrast, our solution runs in time $O(X + Z)$ (§A.8), and works in *streaming* fashion. The key algorithm is CreateTimePrecedenceGraph, given in Figure 6 and proved correct in Appendix A (Lemma 2). The algorithm tracks a "frontier": the set of latest, mutually concurrent requests. Every new arrival descends from all members of the frontier. Once a request leaves, it evicts all of its parents from the frontier. This algorithm may be of independent interest; for example, it could be used to accelerate prior work [14, 51].

Overall, the algorithms in Figures 5 and 6 cost $O(X+Y+Z)$ time and $O(Y)$ space (§A.8), with good constants (Fig. 9; §5.2); here, $Y$ is the number of object operations in the logs.

# 4 A built system: OROCHI

The prior two sections described the Efficient Server Audit Problem, and how it can be solved with SSCO. This section applies the model to an example system that we built.

Consider again Dana, who wishes to verify execution of a SQL-backed PHP web application running on AWS. In this context, the *program* is a PHP application (and the separate PHP scripts are subroutines). The *executor* is the entire remote stack, from the hardware to the hypervisor and all the way up to and including the PHP runtime; we often call the executor just the *server*. The *requests* and *responses* are the HTTP requests and responses that flow in and out of the application. The *collector* is a middlebox at the edge of Dana's company, and is placed to inspect and capture end-clients' requests and the responses that they receive. An *object* can be a SQL database, per-client data that persists across requests, or other external state accessed by the application.

We can apply SSCO to this context, if we:

- Develop a record-replay system for PHP in which replay is batched according to SIMD-on-demand (§3.1).
- Define a set of object types that (a) abstract PHP state constructs (session data, databases, etc.) and (b) obey the semantics in SSCO (§3.2). Each object type requires adapting simulate-and-check (§3.3) and, possibly, modifying the application to respect the interfaces of these objects.
- Incorporate the capture (and ideally validation) of certain sources of non-determinism, such as PHP built-ins.

The above items represent the main work of our system, OROCHI. We describe the details in Sections 4.3–4.7.

## 4.1 Applicability of OROCHI, theory vs. practice

OROCHI is relevant in scenarios besides Dana's. As an example, Pat the Principal runs a public-facing web application on local hardware and is worried about compromise of the server, but trusts a middlebox in front of the server to collect the trace. We describe other scenarios later (§7).

OROCHI is implemented for PHP-based HTTP applications but in principle generalizes to other web standards. Also, OROCHI verifies an application's interactions with its *clients*; verifying communication with external services requires additional mechanism (§5.5). Ultimately, OROCHI is geared to applications with few such interactions. This is certainly restrictive, but there is a useful class within scope: the LAMP [3] stack. The canonical LAMP application is a PHP front-end to a database, for example a wiki or bug database.

The model in Sections 2 and 3 was very general and abstracted away certain considerations that are relevant in OROCHI's setting. We describe these below:

*Persistent objects.* The verifier needs the server's objects as they were at the beginning of the audited period. If audit periods are contiguous, then the verifier in OROCHI produces the required state during the previous audit (§4.5).

*Server-client collusion.* In Section 2, we made no assumptions about the server and clients. Here, however, we assume that the server cannot cause end-clients to issue spurious requests; otherwise, the server might be able to "legally" insert events into history. This assumption fits Dana's situation though is admittedly shakier in Pat's.

*Differences in stack versions.* The verifier's and server's stacks need not be the same. However, it is conceivable that different versions could cause the verifier to erroneously reject a well-behaved server (the inverse error does not arise: validity is defined by the verifier's re-execution). If the verifier wanted to eliminate this risk, it could run a stack with precise functional equivalence to the server's. Another option is to obtain the server-side stack in the event of a divergent re-execution, so as to exonerate the server if warranted.

*Modifications by the network.* Responses modified en route to the collector appear to OROCHI to be the server's responses; modifications between the collector and end-clients—a real concern in Pat's scenario, given that ISPs have hosted ad-inserting middleboxes [31, 91]—can be addressed by Web Tripwires (WT) [70], which are complementary to OROCHI.

## 4.2 Some basics of PHP

PHP [5] is a high-level language. When a PHP script is run by a web server, the components of HTTP requests are materialized as program variables. For example, if the end-user submits `http://www.site.org/s.php?a=7`, then the web server invokes a PHP runtime that executes `s.php`; within the script `s.php`, `$_GET['a']` evaluates to 7.

The data types are *primitive* (int, double, bool, string); *container* (arrays, objects); *reference*; *class*; *resource* (an abstraction of an external resource, such as a connection to a database system); and *callables* (closures, anonymous functions, callable objects).

The PHP runtime translates each program line to byte code: one or more virtual machine (VM) instructions, together with their operands. (Some PHP implementations, such as HHVM, support JIT, though OROCHI's verifier does not support this mode.) Besides running PHP code, the PHP VM can call built-in functions, written in C/C++.

## 4.3 SIMD-on-demand execution in OROCHI

The server and verifier run modified PHP runtimes. The server's maintains an incremental digest for each execution. When the program reaches a branch, this runtime updates the digest based on the type of the branch (jump, switch, or iteration) and the location to which the program jumps. The digest thereby identifies the control flow, and the server records it.

The verifier's PHP runtime is called *acc-PHP*; it performs SIMD-on-demand execution (§3.1), as we describe below.

Acc-PHP works at the VM level, though in our examples and description below, we will be loose and often refer to the original source. Acc-PHP broadens the set of PHP types to include *multivalue* versions of the basic types. For example, a multivalue int can be thought of as a vector of ints. A container's cells can hold multivalues; and a container can itself be a multivalue. Analogously, a reference can name a multivalue; and a reference can itself be a multivalue, in which case each of the references in the vector is logically distinct. A variable that is not a multivalue is called a *univalue*.

All requests in a control flow group invoke the same PHP script *s*. At the beginning of re-executing a control flow group, acc-PHP sets the input variables in *s* to multivalues, based on the inputs in the trace. Roughly speaking, instructions with univalue operands produce univalues, and instructions with multivalue operands produce multivalues. But when acc-PHP produces a multivalue whose components are identical, reflecting a variable that is the same across executions, acc-PHP collapses it down to a univalue; this is crucial to deduplication (§5.2). A collapse is all or nothing: every multivalue has cardinality equal to the number of requests being re-executed.

**Primitive types.** When the operands of an instruction or function are primitive multivalues, acc-PHP executes that instruction or function componentwise. Also, if there are mixed multivalue and univalue operands, acc-PHP performs scalar expansion (as in Matlab, etc.): it creates a multivalue, all of whose components are equal to the original univalue. As an example, consider:

```
1    $sum = $_GET['x'] + $_GET['y'];
2    $larger = max ($sum, $_GET['z']);
3    $odd = ($larger % 2) ? "True" : "False";
4    echo $odd;

r1: /prog.php?x=1&y=3&z=10
r2: /prog.php?x=2&y=4&z=10
```

There are two requests: `r1` and `r2`. Each has three inputs: `x`, `y`, and `z`, which are materialized in the program as `$_GET['x']`, etc. Acc-PHP represents these inputs as multivalues: `$_GET['x']` evaluates to $[1, 2]$, and `$_GET['y']` evaluates to $[3, 4]$. In line 1, both operands of + are multivalues, and `$sum` receives the elementwise sum: $[4, 6]$. In line 2, `$larger` receives $[10, 10]$, and acc-PHP merges the multivalue to make it a univalue. As a result, lines 3 and 4 execute once, rather than once for each request.

A multivalue can comprise different types. For example, in two requests that took the same code path, a program variable was an int in one request and a float in the other. Our acc-PHP implementation handles an int-and-float mixture. However, if acc-PHP encounters a different mixture, it retries, by separately re-executing the requests in sequence.

**Containers.** We use the example of a "set" on an object: `$obj->$key = $val`. Acc-PHP handles "gets" similarly, and likewise other containers (arrays, arrays of arrays, etc.).

Assume first that `$obj` is a multivalue. If either of `$key` and `$val` are univalues, acc-PHP performs scalar expansion to create a multivalue for `$key` and `$val`. Then, acc-PHP assigns the *i*th component of `$val` to the property named by the *i*th component of `$key` in the *i*th object in `$obj`.

Now, if `$obj` is a univalue and `$key` is a multivalue, acc-PHP expands the `$obj` into a multivalue, performs scalar expansion on `$val` (if a univalue), and then proceeds as in the preceding paragraph. The reason for the expansion is that in the original executions, the objects were no longer equivalent.

When `$obj` and `$key` are univalues, and `$val` is a multivalue, acc-PHP assigns `$val` to the given object's given property. This is similar to the way that acc-PHP set up `$_GET['a']` as a multivalue in the example above.

**Built-in functions.** For acc-PHP's re-execution to be correct, PHP's built-in functions (§4.2) would need to be extended to understand multivalues, perform scalar expansion as needed, etc. But there are thousands of built-in functions.

To avoid modifying them all, acc-PHP does the following. When invoking a built-in function, it checks whether any of the arguments are multivalues (if the function is a built-in method, it also checks whether `$this` is a multivalue). If so, acc-PHP splits the multivalue argument into a set of univalues; assume for ease of exposition that there is only one such multivalue argument. Acc-PHP then clones the environment (argument list, function frame); performs a deep copy of any objects referenced by any of the arguments; and executes the function, once for each univalue. Finally, acc-PHP returns the separate function results as a multivalue and maintains the object copies as multivalues. The reason for the deep copy is that the built-in function could have modified the object differently in the original executions.

**Global variables.** There are two cases to handle. First, if a multi-invoked built-in (as above) modifies a global and if the global is a univalue, acc-PHP dynamically expands it to a multivalue. Second, a global can be modified by PHP code, implicitly. For example, referencing a nonexistent property from an object causes invocation of a PHP function, known as a magic method [4], which could modify a global. Acc-PHP detects this case, and expands the global into a multivalue.

### 4.4 Concurrency and shared objects in OROCHI

SSCO's concurrency model (§3.2) fits PHP-based applications, which commonly have concurrent threads, each handling a single end-client request sequentially. OROCHI supports several objects that obey SSCO's required semantics (§3.2) and that abstract key PHP programming constructs:

- *Registers*, with atomic semantics [55]. These work well for modeling per-user persistent state, known as "session data." Specifically, PHP applications index per-user state by browser cookie (this is the "name" of the register) and materialize the state in a program variable. Constructing this variable is the "read" operation; a "write" is performed

- by PHP code, or by the runtime at the end of a request.
- *Key-value stores*, exposing a single-key get/set interface, with linearizable semantics [46]. This models various PHP structures that provide shared memory to requests: the Alternative PHP Cache (APC), etc.
- *SQL databases*, which support single-query statements and multi-query transactions. To make a SQL database behave as one atomic object, we impose two restrictions. First, the database's isolation level must be strict serializability [22, 63].[1] Second, a multi-statement transaction cannot enclose *other* object operations (such as a nested transaction).

The first DB restriction can be met by configuration, as many DBMSes provide strict serializability as an option. However, this isolation level sacrifices some concurrency compared to, say, MySQL's default [7]. The second DB restriction sometimes necessitates minor code changes, depending on the application (§5.4).

To adapt simulate-and-check to an object type, OROCHI must first collect an operation log (§3.3). To that end, some entity (this step is untrusted) wraps relevant PHP statements, to invoke a recording library. Second, OROCHI's verifier needs a mechanism for efficiently re-executing operations on the object. We showed the solution for registers in §3.3. But that technique would not be efficient for databases or key-value stores: to re-execute a DB "select" query, for example, could require going backward through the entire log.

### 4.5 Adapting simulate-and-check to databases

Given a database object $d$—OROCHI handles key-value stores similarly—the verifier performs a *versioned redo* pass over $OL_d$ at the beginning of the audit: it issues every transaction to a *versioned database* [27, 41, 62, 81], setting the version to be the sequence number in $OL_d$. During re-execution, the verifier handles a "write" query (UPDATE, etc.) by checking that the program-generated SQL matches the opcontents field in the corresponding log entry. The verifier handles "read" queries (SELECT, etc.) by issuing the SQL to the versioned DB, specifying the version to be the log sequence number of the current operation. The foregoing corresponds to an additional step in SSCO_AUDIT and further cases in SimOp (Figure 3); the augmented algorithms are in Appendix A.

As an optimization, OROCHI applies *read query deduplication*. If two SELECT queries $P$ and $Q$ are lexically identical and if the parts of the DB covered by $P$ and $Q$ do not change between the redo of $P$ and $Q$, then it suffices to issue the query once during re-execution. To exploit this fact, the verifier, during re-execution, clusters all queries in a control flow group and sorts each cluster by version number. Within a cluster, it de-duplicates queries $P$ and $Q$ if the tables that $P$ and $Q$ touch were not modified between $P$'s and $Q$'s versions.

---

[1] Confusingly, our required atomicity is, in the context of ACID databases, not the "A" but the kind of "I" (isolation); see Bailis [17] for an untangling.

To speed the versioned redo pass, the verifier directs update queries to an *in-memory* versioned database $M$, which acts as a buffer in front of the audit-time versioned database $V$. When the log is fully consumed, the verifier migrates the final state of $M$ to $V$ using a small number of transactions: the verifier dumps each table in $M$ as a single SQL update statement that, when issued to $V$, reproduces the table. The migration could also happen when $M$ reaches a memory limit (although we do not implement this). This would require subsequently re-populating $M$ by reading records from $V$.

### 4.6 Non-determinism

OROCHI includes non-determinism that is not part of the SSCO model: non-deterministic PHP built-ins (time, getpid, etc.), non-determinism in a database (e.g., auto increment ids), and whether a given transaction aborts.

Replay systems commonly record non-determinism during online execution and then, during replay, supply the recorded information in response to a non-deterministic call (see §6.3 for references). OROCHI does this too. Specifically, OROCHI adds a fourth report type (§3): non-deterministic information, such as the return values of certain PHP built-in invocations. The server collects these reports by wrapping the relevant PHP statements (as in §4.4).

But, because reports are untrusted, OROCHI's verifier also *checks* the reported non-determinism against expected behavior. For example, the verifier checks that queries about time are monotonically increasing and that the process id is constant within requests. For random numbers, the application could seed a pseudorandom number generator, and the seed would be the non-deterministic report, though we have not implemented this.

Unfortunately, we cannot give rigorous guarantees about the efficacy of these checks, as our definitions and proofs (Appendix A) do not capture this kind of non-determinism. This is disappointing, but the issue seems fundamental, unless we pull the semantics of PHP into our proofs. Furthermore, this issue exists in all systems that "check" an untrusted lower layer's return values for validity [15, 18, 29, 47, 95].

Beyond that, the server gets discretion over the thread schedule, which is a kind of non-determinism, albeit one that is captured by our definitions and proofs (Appendix A). As an example, if the web service performs a lottery, the server could delay responding to a collection of requests, invoke the random number library, choose which request wins, and then arrange the reports and responses accordingly.

### 4.7 Implementation details

Figure 7 depicts the main components of OROCHI.

A rewrite tool performs required PHP application modifications: inserting wrappers (§4.4, §4.6), and adding hooks to record control flow digests and maximum operation number. Given some engineering, this rewriting can be fully automatic;

| OROCHI component | Base | LOC written/changed |
|---|---|---|
| Server PHP (§4.3) | HHVM [8] | 400 lines of C++ |
| Acc-PHP (§4.3–§4.6) | HHVM [8] | 13k lines of C++ |
| Record library (§4.4, §4.6) | N/A | 1.6k lines of PHP |
| DB logging (§4.4) | MySQL | 320 lines of C++ |
| In-memory versioned DB (§4.5) | SQLite | 1.8k lines of C++ |
| Other audit logic (§3, §4) | N/A | 2.5k lines of C++/PHP/Bash |
| Rewriting tool (§4.7) | N/A | 470 lines of Python, Bash |

**Figure 7**—OROCHI's software components.

our implementation sometimes needs manual help.

To log DB operations (§4.4), the server's PHP runtime passes (*rid*, *opnum*) in the comment field of a SQL query; our code in MySQL (v5.6) assigns a unique sequence number to the query (or transaction), necessitating minor synchronization. Each DB connection locally logs its queries in *sub-logs*; later, a stitching daemon merges these sub-logs to create the database operation log.

OROCHI's versioned DB implementation (§4.5) borrows Warp's [27] schema, and uses the same query rewriting technique (see also §6.2). We implemented OROCHI's audit-time key-value store as a new component (in acc-PHP) to provide a versioned put/get interface.

Acc-PHP has several implementation limitations. One is the limited handling of mixed types, mentioned earlier (§4.3); another is that an object that points to itself (such as $a->b->a) is not recognized as such, if the object is a multivalue. When acc-PHP encounters such cases, it re-executes requests separately. In addition, acc-PHP runs with a maximum number of requests in a control flow group (3,000 in our implementation); this is because the memory consumed by larger sizes would cause thrashing and slow down re-execution.

In OROCHI, the server must be drained prior to an audit, but this is not fundamental; natural extensions of the algorithms would handle prefixes or suffixes of requests' executions.

## 5 Evaluation of OROCHI

This section answers the following questions:

- How do OROCHI's verifier speedup and server overhead compare to a baseline of simple re-execution? (§5.1)
- What are the sources of acceleration? (§5.2)
- What is the "price of verifiability", meaning OROCHI's costs compared to the legacy configuration? (§5.3)
- What kinds of web applications work with OROCHI? (§5.4)

**Applications and workloads.** We answer the first two questions with experiments, which use three applications: MediaWiki (a wiki used by Wikipedia and others), phpBB (an open source bulletin board), and HotCRP (a conference review application). These applications stress different workloads. Also, MediaWiki and phpBB are in common use, and HotCRP has become a reference point for systems security publications that deal with PHP-based web applications [27, 53, 68, 69, 73, 93]. Indeed, MediaWiki and HotCRP are the applications evaluated by Poirot [53] (§6.3). Our experimental workloads are as follows:

*MediaWiki* (v1.26.2). Our workload is derived from a 2007 Wikipedia trace, which we downsampled to 20,000 requests to 200 pages, while retaining its Zipf distribution ($\beta = 0.53$) [85]. We used a 10 year-old trace because we were unable to find something more recent; we downsampled because the original has billions of requests to millions of pages, which is too large for our testbed (on the other hand, smaller workloads produce fewer batching opportunities so are pessimistic to OROCHI).

*phpBB* (v3.2.0). On September 21, 2017, we pulled posts created over the preceding week from a real-world phpBB instance: CentOS [2]. We chose the most popular topic. There were 63 posts, tens to thousands of views per post, and zero to tens of replies per post. We assume that the ratio of page views from registered users (who log in) to guests (who do not) is 1:40, based on sampling reports from the forum (4–9 registered users and 200–414 guests were online). We create 83 users (the number of distinct users in the posts) to view and reply to the posts. The workload contains 30k requests.

*HotCRP*. We build a workload from 269 papers, 58 reviewers, and 820 reviews, with average review length of 3625 characters; the numbers are from SIGCOMM 2009 [9, 64]. We impose synthetic parameters: one registered author submits one valid paper, with a number of updates distributed uniformly from 1 to 20; each paper gets 3 reviews; each reviewer submits two versions of each review; and each reviewer views 100 pages. In all, there are 52k requests.

As detailed later (§5.4), we made relatively small modifications to these applications. A limitation of our investigation is that all modeled clients use the same browser; however, our preliminary investigation indicates that PHP control flow is insensitive to browser details.

**Setup and measurement.** Our testbed comprises two machines connected to a switch. Each machine has a 3.3GHz Intel i5-6600 (4-core) CPU with 16GB memory and a 250GB SSD, and runs Ubuntu 14.04. One of the machines alternates between the roles of server (running Nginx 1.4.6) and verifier; the other generates load. We measure CPU costs from Linux's /proc. We measure throughput and latency at the client.

### 5.1 OROCHI versus the baseline

**What is the baseline?** We want to compare OROCHI to a system that audits comprehensively without trusting reports. A possibility is probabilistic proofs [21, 24, 33, 66, 76, 89], but they cannot handle our workloads, so we would have to estimate, and the estimates would yield outlandish speedups for OROCHI (over $10^6 \times$). Another option is untrusted full-machine replay, as in AVM [44]. However, AVM's imple-mentation supports only single-core servers, and handling untrusted reports *and* concurrency in VM replay might require research (§7).

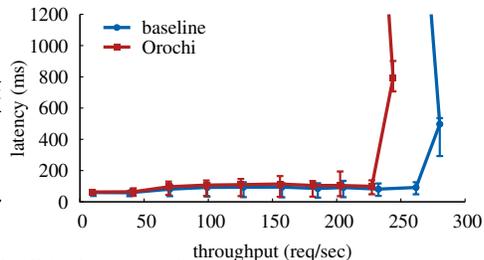

| App | audit speedup | server CPU overhead | avg request | reports (per request) baseline | reports (per request) OROCHI | reports (per request) OROCHI ovhd | DB overhead temp | DB overhead permanent |
|---|---|---|---|---|---|---|---|---|
| MediaWiki | 10.9× | 4.7% | 7.1KB | 0.8KB | 1.7KB | 11.4% | 1.0× | 1× |
| phpBB | 5.6× | 8.6% | 5.7KB | 0.1KB | 0.3KB | 2.7% | 1.7× | 1× |
| HotCRP | 6.2× | 5.9% | 3.2KB | 0.0KB | 0.4KB | 10.9% | 1.5× | 1× |

**Figure 8**—OROCHI compared to simple re-execution (§5.1). *Left table*: "Audit speedup" is the ratio of audit-time CPU costs, assuming (conservatively) that auditing in simple re-execution is the same cost as serving the legacy application, and (perhaps optimistically) that simple re-execution and OROCHI are given HTTP requests and responses from the trace collector. "Server CPU overhead" is the CPU cost added by OROCHI, conservatively assuming that the baseline imposes no server CPU costs. The reports are compressed (OROCHI's overheads include the CPU cost of compression/decompression; the baseline is not charged for this). "OROCHI ovhd" in those columns is the ratio of (the trace plus OROCHI's reports) to (the trace plus the baseline's reports). "Temp" DB overhead refers to the ratio of the size of the on-disk versioned DB (§4.5) to the size of a non-versioned DB. *Right graph:* Latency vs. server throughput for phpBB (the other two workloads are similar). Points are 90th (bars are 50th and 99th) percentile latency for a given request rate, generated by a Poisson process. The depicted data are the medians of their respective statistics over 5 runs.

Instead, we evaluate against a baseline that is less expensive than both of these approaches, and hence is pessimistic to OROCHI: the legacy application (without OROCHI), which can be seen as a lower bound on hypothetical *simple re-execution*.

We capture this baseline's *audit-time CPU cost* by measuring the legacy server CPU costs; in reality, an audit not designed for acceleration would likely proceed more slowly. We assume this baseline has no *server CPU overhead*; in reality, the baseline would have some overhead. We capture the baseline's *report size* with OROCHI's non-deterministic reports (§4.6), because record-replay systems need non-deterministic advice; in reality, the baseline would likely need additional reports to reconstruct the thread schedule. Finally, we assume that the baseline tolerates arbitrary database configurations (unlike OROCHI; §4.4), although we assume that the baseline needs to reconstruct the database (as in OROCHI).

**Comparison.** Figure 8 compares OROCHI to the aforementioned baseline. At a high level, OROCHI accelerates the audit compared to the baseline (we delve into this in §5.2) but introduces some server CPU cost, with some degradation in throughput, and minor degradation in latency.

The throughput reductions are respectively 13.0%, 11.1% and 17.8% for phpBB, MediaWiki, and HotCRP. The throughput comparison includes the effect of requiring strict serializability (§4.4), because the baseline's databases are configured with MySQL's default isolation level (repeatable read).

The report overhead depends on the frequency of object operations (§4.4) and non-deterministic calls (§4.6). Still, the report size is generally a small fraction of the size of the trace, as is OROCHI's "report overhead" versus the baseline. OROCHI's audit-time DB storage requirement is higher than the baseline's, because of versioning (§4.5), but after the audit, OROCHI needs only the "latest" state.

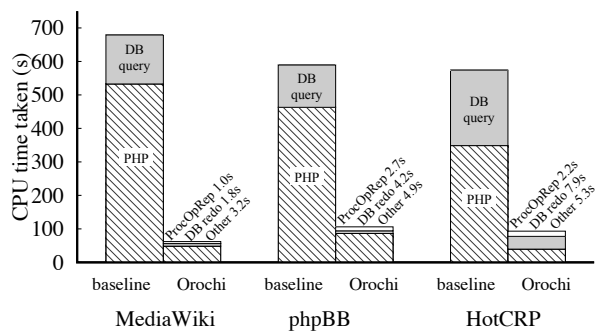

**Figure 9**—Decomposition of audit-time CPU costs. "PHP" (in OROCHI) is the time to perform SIMD-on-demand execution (§3.1,§4.3) and simulate-and-check (§3.3,§4.4). "DB query" is the time spent on DB queries during re-execution (§4.5). "ProcOpRep" is the time to execute the logic in Figures 5 and 6. "DB redo" is the time to reconstruct the versioned storage (§4.5). "Other" includes miscellaneous costs such as initializing inputs as multivalues, output comparison, etc.

### 5.2 A closer look at acceleration

Figure 9 decomposes the audit-time CPU costs. The "DB query" portion illustrates query deduplication (§4.5). Without this technique, every DB operation would have to be re-issued during re-execution. (OROCHI's verifier re-issues every register and key-value operation, but these are inexpensive.) Query deduplication is more effective when the workload is read-dominated, as in our MediaWiki experiment.

We now investigate the sources of PHP acceleration; we wish to know the costs and benefits of univalent and multivalent instructions (§3.1, §4.3). We divide the 100+ PHP byte code instructions into 10 categories (arithmetic, container, control flow, etc.); choose category representatives;

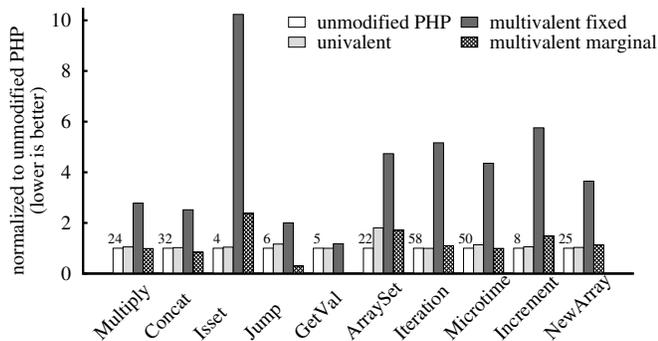

**Figure 10**—Cost of various instructions in unmodified PHP and acc-PHP (§4.3). Execution times are normalized clusterwise to unmodified PHP, for which the absolute time is given (in $\mu$s). See text for interpretation.

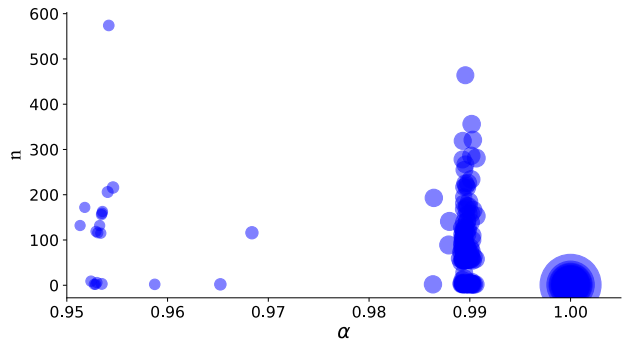

**Figure 11**—Characteristics of control flow groups in the MediaWiki workload. Each bubble is a control flow group; the center of a bubble gives the group's $n$ (number of requests in the group) and $\alpha$ (proportion of univalent instructions); the size of a bubble is proportional to $\ell$ (number of instructions in the group). This workload has 527 total groups (bubbles), 237 groups with $n > 1$, and 200 unique URLs. All groups have $\alpha > 0.95$; only the occupied portion of the $x$-axis is depicted.

and run a microbenchmark that performs $10^7$ invocations of the instruction and computes the average cost. We run each microbenchmark against unmodified PHP, acc-PHP with univalent instructions, and acc-PHP with multivalent instructions; we decompose the latter into marginal cost (the cost of an additional request in the group) and fixed cost (the cost if acc-PHP were maintaining a multivalue with zero requests).

Figure 10 depicts the results. The fixed cost of multivalent instructions is high, and the marginal cost is sometimes worse than the unmodified baseline. In general, multivalent execution is *worse* than simply executing the instruction $n$ times![2] So how does OROCHI accelerate? We hypothesize that (i) many requests share control flow, and (ii) within a shared control flow group, the vast majority of instructions are executed univalently. If this holds, then the gain of SIMD-on-demand execution comes not from the "SIMD" part but rather from the "on demand" part: the opportunistic collapsing of multivalues enables a lot of deduplication.

To confirm the hypothesis, we analyze all of the control flow groups in our workloads. Each group $c$ is assigned a triple $(n_c, \alpha_c, \ell_c)$, where $n_c$ is the number of requests in the group, $\alpha_c$ is the proportion of univalent instructions in that group, and $\ell_c$ is the number of instructions in the group. (Note that if $n_c = 1$, then $\alpha_c = 1.0$.) Figure 11 depicts these triples for the MediaWiki workload. There are many groups with high $n_c$, and most groups have very high $\alpha_c$ (the same holds for the other two workloads), confirming our hypothesis. Something else to note is a slight negative correlation between $n_c$ and $\alpha_c$ within a workload, which is not ideal for OROCHI.

### 5.3 The price of verifiability

We now take stock of OROCHI's total overhead by comparing OROCHI to the *legacy configuration*. OROCHI introduces a

---

[2] One might wonder: would it be better to batch by control flow *and* identical inputs? No; that approach still produces multivalent executions because of shared object reads and non-determinism, and the batch sizes are smaller.

modest cost to the server: 4.7%–8.6% CPU overhead (Figure 8) and temporary storage for trace and reports. But the main price of verifiability is the verifier's resources:

*CPU.* Since the verifier's audit-time CPU costs are between 1/5.6 and 1/10.9 those of the server's costs (per §5.1, Figure 8), OROCHI requires that the verifier have 9.1%–18.0% of the CPU capacity that the server does.

*Storage.* The verifier has to store the database between audits, so the verifier effectively maintains a copy of the database. During the audit, the verifier also stores the trace, reports, and additional DB state (the versioning information).

*Network.* The verifier receives the trace and reports over the network. Note that in the Dana (§1) and Pat (§4.1) scenarios, the principal is already paying (on behalf of clients or the server, respectively) to send requests and responses over the wide area network—which likely swamps the cost of sending the same data to the verifier over a local network.

### 5.4 Compatibility

We performed an informal survey of popular PHP applications to understand the effect of OROCHI's two major compatibility restrictions: the non-verification of interactions with other applications (§4.1) and the non-nesting of object operations inside DB transactions (§4.4).

We sorted GitHub Trending by stars in decreasing order, filtered for PHP applications (filtering out projects that are libraries or plugins), and chose the top 10: Wordpress, Piwik, Cachet, October, Paperwork, Magento2, Pagekit, Lychee, Opencart, and Drupal. We inspected the code (and its configuration and documentation), ran it, and logged object operations. For eight of them, the sole external service is email; the other two (Magento2 and Opencart) additionally interact with a payment server. Also, all but Drupal and Octo-ber are consistent with the DB requirement.

This study does *not* imply that OROCHI runs with these applications out of the box. It generally takes some adjustment to fit an application to OROCHI, as we outline below.

MediaWiki does not obey the DB requirement. We modified it so that requests read in the relevant APC keys (which we obtain through static inspection plus a dynamic list of needed keys, itself stored in the APC), execute against a local cache of those keys, and flush them back to the APC. This gives up some consistency in the APC, but MediaWiki anyway assumes that the APC is providing loose consistency. We made several other minor modifications to MediaWiki; for example, changing an absolute path (stored in the database) to a relative one. In all, we modified 346 lines of MediaWiki (of 410k total and 74k invoked in our experiments).

We also modified phpBB (270 lines, of 300k total and 44k invoked), to address a SQL parsing difference between the actual database (§4.4) and the in-memory one (§4.5) and to create more audit-time acceleration opportunities (by reducing the frequency of updates to login times and page view counters). We modify HotCRP (67 lines, of 53k total and 37k invoked), mainly to rewrite SELECT * FROM queries to request individual columns; the original would fetch the begin/end timestamp columns in the versioned DB (§4.5, §4.7).

### 5.5 Discussion and limitations of OROCHI

Below we summarize OROCHI and discuss its limitations.

*Guarantees.* OROCHI is based on SSCO, which has provable properties. However, OROCHI does not provide SSCO's idealized Soundness guarantee (§2), because of the leeway discussed earlier (§4.6). And observable differences in the verifier's and server's stacks (§4.1) would make OROCHI fall short of SSCO's idealized Completeness guarantee.

*Performance and price.* Relative to a pessimistic baseline, OROCHI's verifier accelerates by factors between 5.6–10.9× in our experiments, and server overhead is below 10% (§5.1). The CPU costs introduced by OROCHI are small, compared to what one sometimes sees in secure systems research; one reason is that OROCHI is not based on cryptography. And while the biggest percentage cost for the verifier is storage (because the verifier has to duplicate it; §5.3), storage is generally inexpensive in dollar terms.

*Compatibility and usability.* On the one hand, OROCHI is limited to a class of applications, as discussed (§4.1, §5.4). On the other hand, the applications in our experiments—which were largely chosen by following prior work (discussed early in §5)—did not require much modification (§5.4). Best of all, OROCHI is fully compatible with today's *infrastructure*: it works with today's end-clients and cloud offerings as-is.

Of course, OROCHI would benefit from extensions. All of the applications we surveyed make requests of an email server (§5.4). We could verify those requests—but not the email server itself; that is future work—with a modest addition to OROCHI, namely treating external requests as another kind of response. This would require capturing the requests themselves; that could be done, in Pat's scenario (§4.1), by the trace collector or, in Dana's scenario (§1), by redirecting email to a trusted proxy on the verifier.

Another extension is adding a *file* abstraction to our three object types (§4.4). This isn't crucial—many applications, including five of the 10 in our survey (§5.4), can be configured to use alternatives such as a key-value store—but some deployers might prefer a file system back-end. Another extension is filtering large objects from the trace, before it is delivered to the verifier. A possible solution is to leverage browser support for Resource Integrity: the verifier would check that the correct *digest* was supplied to the browser, leaving the actual object check to the browser. Other future work is HTTPS; one option is for the server to record non-deterministic cryptographic input, and the verifier uses it to recover the plaintext stream.

A more fundamental limitation is that if OROCHI's verifier does not have a trace from a period (for example, before OROCHI was deployed on a given server), then OROCHI can verify only by getting the pre-OROCHI collection of objects from the server (requiring a large download) and treating those objects as the true initial state (requiring trust).

## 6 Related work

### 6.1 Efficient execution integrity

Efficient execution integrity—giving some *principal* confidence that an *executor*'s outputs are consistent with an expected *program*, without requiring the principal to re-execute the program—is a broad topic. The Efficient Server Audit Problem (§2) combines for the first time: (1) no assumptions about the executor (though our verifier gets a trace of requests/responses), (2) a concurrent executor, and (3) a requirement of scaling to real applications, including legacy ones.

A classic solution is Byzantine replication [26]; the principal needs no verification algorithm but assumes that a super-majority of nodes operates fault-free. Another classic technique is attestation: proving to the principal that the executor runs the expected software. This includes TPM-based approaches [28, 45, 59, 60, 67, 72, 75, 80] and systems [15, 18, 49, 74, 78] built on SGX hardware [50]. But attesting to a (possibly vulnerable) stack does not guarantee the execution integrity of the program atop that stack. Using SGX, we can place the program in its own *enclave*, but it is difficult to rigorously establish that the checks performed by the in-enclave code on the out-enclave code [15, 18, 78] comprehensively detect deviations from expected behavior (though see [79]).

EVE [51] spot-checks for storage consistency violations but assumes correct application execution. Like OROCHI (§4), Verena [52] targets web applications; it doesn't require a trace but does assume a trusted hash server. Verena's techniques are built on authenticated data structures with a restricted API; it does not support general-purpose or legacy web applications.

Execution integrity has long been studied by theorists [16, 38, 39, 43, 61], and these ideas have been refined and implemented [20, 33, 66, 76] (see [89] for a survey and [13, 19, 88, 96] for recent developments). This theory makes no assumptions about the executor or the workload. But it doesn't handle a concurrent executor. Also, because these works generally represent programs as static circuits in which state operations exhaust a very limited "gate budget", and because the executor's overhead is generally at least six orders of magnitude, they are for now unsuited to legacy web applications.

### 6.2 Related techniques

**Computation deduplication.** Delta execution [84] validates patches in C programs by running the patched and unpatched code together; it attempts to execute only the deltas, using copy-on-write fork and merging. In incremental computation (see data-triggered threads [83], iThreads [23], UNIC [82], and citations therein), a program runs once and, when the input changes, only the dependent parts rerun. In contrast, SIMD-on-demand (§3.1) works at a higher level of abstraction; this exposes deduplication opportunities [53] and allows the verifier and executor to run separate implementations of the same logical program (§7). Also, SIMD-on-demand re-executes multiple requests *simultaneously*, which composes easily with query deduplication (§4.5).

**Consistency testing.** Anderson et al. [14] give an algorithm that checks whether a trace of operations on a key-value store obeys *register* semantics [55]; the algorithm builds a graph with time and precedence edges, and checks whether it is acyclic. (See EVE [51] for a related algorithm, and Gibbons-Korach [40] and others [42, 90] for consistency testing in general; see also [10, 11, 77] for related algorithms that analyze programs for memory consistency.) The time edges and cycle detection in SSCO (Fig. 5) are reminiscent of Anderson et al.; however, SSCO captures time edges more efficiently, as noted in §3.5. More significantly, SSCO solves a different problem: it validates whether a request trace meets complex *application* semantics (requests are permitted to be intermingled and invoke multiple operations), and reports are untrusted.

**Time travel databases.** As noted (§4.7), OROCHI's versioned DB borrows from Warp [27] (see also [41, 62, 81]). However, OROCHI constructs that DB only during audit, which enables the techniques in §4.5. Also, OROCHI handles multi-statement transactions, which Warp does not implement.

### 6.3 Deterministic record-replay

Record-replay is a mature field [34, 35]. SSCO (with OROCHI as an instantiation) is the first record-replay system to achieve the following combination: (a) the recorder is untrusted (and the replayer has an input/output trace), (b) replay is accelerated versus re-executing, and (c) there are concurrent accesses to shared objects. We elaborate below.

*Untrusted recorder.* In AVM [44], an untrusted hypervisor records alleged network I/O and non-deterministic events. A replayer checks this log against the ground truth network messages and then re-executes, using VM replay [25, 36]. In Ripley [87], a web server re-executes client-side code to determine whether the output matches what the client claimed. In both cases, the replayer does not trust the recorder, but in neither case is re-execution accelerated.

*Accelerated replay.* Poirot [53] accelerates the re-execution of web applications. OROCHI imitates Poirot: we borrow the observation that web applications have repeated control flow and the notion of grouping re-execution accordingly, and we follow some of Poirot's implementation and evaluation choices (§4.7, §5). But there is a crucial distinction. Poirot analyzes patches in application code; its techniques for acceleration (construct templates for each claimed control flow group) and shared objects (replay "reads") fundamentally trust the language runtime and all layers below [53, §2.4].

*Shared objects and concurrency.* We focus on solutions that enable an offline replayer to deterministically re-execute concurrent operations. First, the replayer can be given a thread schedule explicitly [57, 86]. Second, the replayer can be given information to *reconstruct* the thread schedule, for example operation precedence using CREW protocols [30, 37, 54, 56, 94]. Third, the replayer can be given information to *approximately reconstruct* the thread schedule, for example, synchronization precedence or sketches [12, 65, 71].[3] Closest to simulate-and-check (§3.3) is LEAP [48] (see also [92]), which is in the second category: for each shared Java variable, LEAP logs the sequence of thread accesses. But SSCO's logs also contain *operands*. Simulate-and-check relates to record-replay speculation [57]: it is reminiscent of the way that the epoch-parallel processors in DoublePlay [86] check the starting conditions of optimistically executing future splices.

## 7 Future work and conclusion

To recap, we defined a general problem of execution integrity for concurrent servers (§2); exhibited an abstract solution, SSCO, based on new kinds of replay (§3); and described a system, OROCHI, that instantiates SSCO for web applications and runs on today's cloud infrastructure (§4–§5).

OROCHI applies in scenarios beyond those of Dana (§1) and Pat (§4.1). As an example, consider Adrian the AWS User who deploys a *public*-facing web application. To use OROCHI, Adrian needs a trace. Perhaps Adrian trusts AWS to gather it (in which case Adrian's threat model is a remote attacker, not a cloud insider). Or perhaps AWS lets Adrian use an SGX enclave, within which Adrian runs the trace collector together with an HTTPS proxy that holds Adrian's TLS keys; this enforces trace collection and does not trust AWS but does trust the attested trace collection software.

---

[3]DoublePlay [86] and Respec [57] use these techniques but do so online, while searching for a thread schedule to give to an offline replayer.

Another use case is *patch-based auditing*, proposed in Poirot [53] (see also [58, 84]); here, one replays prior requests against patched code to see if the responses are now different. OROCHI can audit the effect of a patch at any layer, not just in PHP code (as in Poirot).

An interesting aspect of SSCO is that the verifier and the server need not run the same program—only the same logic. For example, the executor can be a complex, replicated cloud environment while the verifier can re-execute the logic however it wants, as long as it gets appropriate reports.

Future work is to instantiate SSCO for other web languages, create variations of SSCO for other concurrency models, and extend SSCO to multiple interacting servers. In addition, we think that the techniques of SSCO have wider applicability. For example, a direction to explore is applying query deduplication (§4.5) and simultaneous replay (§3.1) to general-purpose or lower-level record-replay systems.

Another interesting problem is to produce a multiprocessor record-replay system that works in a setting in which reports are untrusted. This problem provides some intuition for our original challenge (§2), so we conclude the paper by pointing out why this problem is difficult.

Suppose that the offline replayer expects an explicit thread schedule from the recorder. Then the recorder could supply a schedule that is inconsistent with any valid execution (for example, a schedule that ignores user-level synchronization). By correlating bogus outputs and a bogus schedule (similar to §3.4), the recorder could cause the replayer to reproduce illegal executions, violating Soundness (§2). If instead the replayer gets sparse constraints from the recorder [12, 65] and expects to synthesize a schedule itself, this would violate Completeness (§2): an adversarial recorder can make the replayer search in vain for a schedule, which means the replayer needs to bound its searching, which means that some *valid* executions will be rejected for lack of search time.

The fundamental difficulty here is that concurrency necessitates reports (for Completeness), but if the reports are untrusted, the replayer could be misled (compromising Soundness). Efficiency adds a further complication. This problem—designing the reports and a procedure that validates them even as it exploits them—was more challenging than we expected.

OROCHI's source code is released at:
https://github.com/naizhengtan/orochi

### Acknowledgments

This paper was substantially improved by detailed comments from Marcos K. Aguilera, Sebastian Angel, Trinabh Gupta, Brad Karp, Jinyang Li, Ioanna Tzialla, Riad Wahby, and the OSDI16 and SOSP17 anonymous reviewers. We thank Chao Xie and Shuai Mu for advice, and our shepherd Jason Flinn for insightful discussions and close readings that improved the paper. Three people were particularly influential: Curtis Li assisted with the development and evaluation of a previous version of OROCHI; Alexis Gallagher provided critical perspective and suggestions on the paper's structure and exposition; and Brad Karp supplied essential wisdom and sanity. This work was supported by NSF grants CNS-1423249 and CNS-1514422, ONR grant N00014-16-1-2154, and AFOSR grant FA9550-15-1-0302.

### References

[1] Amazon Web Services (AWS). https://aws.amazon.com/.
[2] CentOS forum. https://www.centos.org/forums/.
[3] LAMP (software bundle). https://en.wikipedia.org/wiki/LAMP_(software_bundle).
[4] PHP magic methods. http://php.net/manual/en/language.oop5.magic.php.
[5] PHP manual. http://php.net/manual/en/index.php.
[6] SIMD. https://en.wikipedia.org/wiki/SIMD.
[7] Transaction isolation levels. https://dev.mysql.com/doc/refman/5.6/en/innodb-transaction-isolation-levels.html.
[8] A virtual machine designed for executing programs written in Hack and PHP. https://github.com/facebook/hhvm.
[9] The process of ACM Sigcomm 2009. http://www.sigcomm.org/conference-planning/the-process-of-acm-sigcomm-2009.
[10] J. Alglave, D. Kroening, V. Nimal, and D. Poetzl. Don't sit on the fence: A static analysis approach to automatic fence insertion. In *Computer Aided Verification (CAV)*, July 2014.
[11] J. Alglave and L. Maranget. Stability in weak memory models. In *Computer Aided Verification (CAV)*, July 2011.
[12] G. Altekar and I. Stoica. ODR: output-deterministic replay for multicore debugging. In *ACM Symposium on Operating Systems Principles (SOSP)*, Oct. 2009.
[13] S. Ames, C. Hazay, Y. Ishai, and M. Venkitasubramaniam. Ligero: Lightweight sublinear arguments without a trusted setup. In *ACM Conference on Computer and Communications Security (CCS)*, Oct. 2017.
[14] E. Anderson, X. Li, M. A. Shah, J. Tucek, and J. J. Wylie. What consistency does your key-value store *actually* provide? In *USENIX Workshop on Hot Topics in System Dependability (HotDep)*, Oct. 2010. Full version: Technical Report HPL-2010-98, Hewlett-Packard Laboratories, 2010.
[15] S. Arnautov, B. Trach, F. Gregor, T. Knauth, A. Martin, C. Priebe, J. Lind, D. Muthukumaran, D. O'Keeffe, M. L. Stillwell, D. Goltzsche, D. Eyers, R. Kapitza, P. Pietzuch, and C. Fetzer. SCONE: Secure Linux containers with Intel SGX. In *Symposium on Operating Systems Design and Implementation (OSDI)*, Nov. 2016.
[16] L. Babai, L. Fortnow, L. A. Levin, and M. Szegedy. Checking computations in polylogarithmic time. In *ACM Symposium on the Theory of Computing (STOC)*, May 1991.
[17] P. Bailis. Linearizability versus serializability. http://www.bailis.org/blog/linearizability-versus-serializability/, Sept. 2014.
[18] A. Baumann, M. Peinado, and G. Hunt. Shielding applications from an untrusted cloud with Haven. In *Symposium on Operating Systems Design and Implementation (OSDI)*, Oct. 2014.
[19] E. Ben-Sasson, I. Ben-Tov, A. Chiesa, A. Gabizon, D. Genkin, M. Hamilis, E. Pergament, M. Riabzev, M. Silberstein, E. Tromer, and

M. Virza. Computational integrity with a public random string from quasi-linear PCPs. In *Annual International Conference on the Theory and Applications of Cryptographic Techniques (EUROCRYPT)*, Apr. 2017.

[20] E. Ben-Sasson, A. Chiesa, D. Genkin, E. Tromer, and M. Virza. SNARKs for C: Verifying program executions succinctly and in zero knowledge. In *IACR International Cryptology Conference (CRYPTO)*, Aug. 2013.

[21] E. Ben-Sasson, A. Chiesa, E. Tromer, and M. Virza. Succinct non-interactive zero knowledge for a von Neumann architecture. In *USENIX Security*, Aug. 2014.

[22] P. A. Bernstein, D. W. Shipman, and W. S. Wong. Formal aspects of serializability in database concurrency control. *IEEE Transactions on Software Engineering*, SE-5(3), May 1979.

[23] P. Bhatotia, P. Fonseca, U. A. Acar, B. B. Brandenburg, and R. Rodrigues. iThreads: A threading library for parallel incremental computation. In *ACM International Conference on Architectural Support for Programming Languages and Operating Systems (ASPLOS)*, Mar. 2015.

[24] B. Braun, A. J. Feldman, Z. Ren, S. Setty, A. J. Blumberg, and M. Walfish. Verifying computations with state. In *ACM Symposium on Operating Systems Principles (SOSP)*, Nov. 2013.

[25] T. C. Bressoud and F. B. Schneider. Hypervisor-based fault-tolerance. *ACM Transactions on Computer Systems (TOCS)*, 14(1):80–107, 1996.

[26] M. Castro and B. Liskov. Practical Byzantine fault tolerance and proactive recovery. *ACM Transactions on Computer Systems (TOCS)*, 20(4):398–461, 2002.

[27] R. Chandra, T. Kim, M. Shah, N. Narula, and N. Zeldovich. Intrusion recovery for database-backed web applications. In *ACM Symposium on Operating Systems Principles (SOSP)*, Oct. 2011.

[28] C. Chen, P. Maniatis, A. Perrig, A. Vasudevan, and V. Sekar. Towards verifiable resource accounting for outsourced computation. In *ACM Virtual Execution Environments (VEE)*, Mar. 2013.

[29] X. Chen, T. Garfinkel, E. C. Lewis, P. Subrahmanyam, C. A. Waldspurger, D. Boneh, J. Dwoskin, and D. R. K. Ports. Overshadow: a virtualization-based approach to retrofitting protection in commodity operating systems. *ACM International Conference on Architectural Support for Programming Languages and Operating Systems (ASPLOS)*, Mar. 2008.

[30] Y. Chen and H. Chen. Scalable deterministic replay in a parallel full-system emulator. In *ACM Symposium on Principles and Practice of Parallel Programming (PPoPP)*, Feb. 2013.

[31] J. Cheng. NebuAd, ISPs sued over DPI snooping, ad-targeting program. *Ars Technica*, Nov. 2008. https://arstechnica.com/tech-policy/2008/11/nebuad-isps-sued-over-dpi-snooping-ad-targeting-program/.

[32] T. H. Cormen, C. E. Leiserson, R. L. Rivest, and C. Stein. *Introduction to Algorithms, third edition*. The MIT Press, Cambridge, MA, 2009.

[33] G. Cormode, M. Mitzenmacher, and J. Thaler. Practical verified computation with streaming interactive proofs. In *Innovations in Theoretical Computer Science (ITCS)*, Jan. 2012.

[34] F. Cornelis, A. Georges, M. Christiaens, M. Ronsse, T. Ghesquiere, and K. D. Bosschere. A taxonomy of execution replay systems. In *International Conference on Advances in Infrastructure for Electronic Business, Education, Science, Medicine, and Mobile Technologies on the Internet*, 2003.

[35] C. Dionne, M. Feeley, and J. Desbiens. A taxonomy of distributed debuggers based on execution replay. In *Proceedings of the International Conference on Parallel and Distributed Processing Techniques (PDPTA)*, Aug. 1996.

[36] G. W. Dunlap, S. T. King, S. Cinar, M. A. Basrai, and P. M. Chen. ReVirt: Enabling intrusion analysis through virtual-machine logging and replay. In *Symposium on Operating Systems Design and Implementation (OSDI)*, Dec. 2002.

[37] G. W. Dunlap, D. Lucchetti, P. M. Chen, and M. Fetterman. Execution replay for multiprocessor virtual machines. In *ACM Virtual Execution Environments (VEE)*, Mar. 2008.

[38] R. Gennaro, C. Gentry, and B. Parno. Non-interactive verifiable computing: Outsourcing computation to untrusted workers. In *IACR International Cryptology Conference (CRYPTO)*, Aug. 2010.

[39] R. Gennaro, C. Gentry, B. Parno, and M. Raykova. Quadratic span programs and succinct NIZKs without PCPs. In *Annual International Conference on the Theory and Applications of Cryptographic Techniques (EUROCRYPT)*, May 2013.

[40] P. B. Gibbons and E. Korach. Testing shared memories. *SIAM Journal on Computing*, 26(4):1208–1244, 1997.

[41] A. Goel, K. Farhadi, K. Po, and W. Feng. Reconstructing system state for intrusion analysis. *ACM SIGOPS Operating Systems Review*, 42(3):21–28, Apr. 2008.

[42] W. Golab, X. Li, and M. Shah. Analyzing consistency properties for fun and profit. In *PODC*, June 2011.

[43] S. Goldwasser, Y. T. Kalai, and G. N. Rothblum. Delegating computation: Interactive proofs for muggles. *Journal of the ACM*, 62(4):27:1–27:64, Aug. 2015. Prelim version STOC 2008.

[44] A. Haeberlen, P. Aditya, R. Rodrigues, and P. Druschel. Accountable virtual machines. In *Symposium on Operating Systems Design and Implementation (OSDI)*, Oct. 2010.

[45] C. Hawblitzel, J. Howell, J. R. Lorch, A. Narayan, B. Parno, D. Zhang, and B. Zill. Ironclad apps: End-to-end security via automated full-system verification. In *Symposium on Operating Systems Design and Implementation (OSDI)*, Oct. 2014.

[46] M. P. Herlihy and J. M. Wing. Linearizability: A correctness condition for concurrent objects. *ACM Transactions on Programming Languages and Systems (TOPLAS)*, 12(3), July 1990.

[47] O. S. Hofmann, S. Kim, A. M. Dunn, M. Z. Lee, and E. Witchel. InkTag: secure applications on an untrusted operating system. *ACM International Conference on Architectural Support for Programming Languages and Operating Systems (ASPLOS)*, pages 265–278, Mar. 2013.

[48] J. Huang, P. Liu, and C. Zhang. LEAP: The lightweight deterministic multi-processor replay of concurrent Java programs. In *ACM Symposium on the Foundations of Software Engineering (FSE)*, Feb. 2010.

[49] T. Hunt, Z. Zhu, Y. Xu, S. Peter, and E. Witchel. Ryoan: A distributed sandbox for untrusted computation on secret data. In *Symposium on Operating Systems Design and Implementation (OSDI)*, Nov. 2016.

[50] Intel. Intel software guard extensions programming reference. https://software.intel.com/sites/default/files/managed/48/88/329298-002.pdf.

[51] S. Jana and V. Shmatikov. EVE: Verifying correct execution of cloud-hosted web applications. In *USENIX HotCloud Workshop*, June 2011.

[52] N. Karapanos, A. Filios, R. A. Popa, and S. Capkun. Verena: End-to-end integrity protection for web applications. In *IEEE Symposium on Security and Privacy*, May 2016.

[53] T. Kim, R. Chandra, and N. Zeldovich. Efficient patch-based auditing for web applications. In *Symposium on Operating Systems Design and Implementation (OSDI)*, Oct. 2012.

[54] O. Laadan, N. Viennot, and J. Nieh. Transparent, lightweight application execution replay on commodity multiprocessor operating systems. In *SIGMETRICS*, June 2010.

[55] L. Lamport. On interprocess communication, parts I and II. *Distributed Computing*, 1(2):77–101, 1986.

[56] T. J. LeBlanc and J. M. Mellor-Crummey. Debugging parallel programs with Instant Replay. *IEEE Transactions on Computers*, C-36(4):471–482, 1987.

[57] D. Lee, B. Wester, K. Veeraraghavan, S. Narayanasamy, P. M. Chen, and J. Flinn. Respec: Efficient online multiprocessor replay via speculation and external determinism. In *ACM International Conference on Architectural Support for Programming Languages and Operating Systems (ASPLOS)*, Mar. 2010.

[58] M. Maurer and D. Brumley. Tachyon: tandem execution for efficient

live patch testing. *USENIX Security*, pages 617–630, Aug. 2012.

[59] J. M. McCune, Y. Li, N. Qu, Z. Zhou, A. Datta, V. Gligor, and A. Perrig. TrustVisor: Efficient TCB reduction and attestation. In *IEEE Symposium on Security and Privacy*, May 2010.

[60] J. M. McCune, B. J. Parno, A. Perrig, M. K. Reiter, and H. Isozaki. Flicker: An execution infrastructure for TCB minimization. In *European Conference on Computer Systems (EuroSys)*, Apr. 2008.

[61] S. Micali. Computationally sound proofs. *SIAM Journal on Computing*, 30(4):1253–1298, 2000.

[62] Oracle flashback technology. http://www.oracle.com/technetwork/database/features/availability/flashback-overview-082751.html.

[63] C. H. Papadimitriou. The serializability of concurrent database updates. *Journal of the ACM*, 26(4), Oct. 1979.

[64] D. Papagiannaki and L. Rizzio. The ACM SIGCOMM 2009 Technical Program Committee Process. *ACM CCR*, 39(3):43–48, July 2009.

[65] S. Park, Y. Zhou, W. Xiong, Z. Yin, R. Kaushik, K. H. Lee, and S. Lu. PRES: probabilistic replay with execution sketching on multiprocessors. In *ACM Symposium on Operating Systems Principles (SOSP)*, Oct. 2009.

[66] B. Parno, C. Gentry, J. Howell, and M. Raykova. Pinocchio: Nearly practical verifiable computation. In *IEEE Symposium on Security and Privacy*, May 2013.

[67] B. Parno, J. M. McCune, and A. Perrig. *Bootstrapping Trust in Modern Computers*. Springer, 2011.

[68] B. Parno, J. M. McCune, D. Wendlandt, D. G. Andersen, and A. Perrig. CLAMP: Practical prevention of large-scale data leaks. In *IEEE Symposium on Security and Privacy*, May 2009.

[69] R. A. Popa, C. Redfield, N. Zeldovich, and H. Balakrishnan. CryptDB: protecting confidentiality with encrypted query processing. In *ACM Symposium on Operating Systems Principles (SOSP)*, Oct. 2011.

[70] C. Reis, S. D. Gribble, T. Kohno, and N. C. Weaver. Detecting in-flight page changes with Web Tripwires. In *Symposium on Networked Systems Design and Implementation (NSDI)*, Apr. 2008.

[71] M. Ronsse and K. D. Bosschere. RecPlay: a fully integrated practical record/replay system. *ACM Transactions on Computer Systems (TOCS)*, 17(2):133–152, 1999.

[72] R. Sailer, X. Zhang, T. Jaeger, and L. van Doorn. Design and implementation of a TCG-based integrity measurement architecture. In *USENIX Security*, Aug. 2004.

[73] D. Schultz and B. Liskov. Ifdb: decentralized information flow control for databases. In *European Conference on Computer Systems (EuroSys)*, Apr. 2013.

[74] F. Schuster, M. Costa, C. Fournet, C. Gkantsidis, M. Peinado, G. Mainar-Ruiz, and M. Russinovich. VC3: Trustworthy data analytics in the cloud using SGX. In *IEEE Symposium on Security and Privacy*, May 2015.

[75] A. Seshadri, M. Luk, E. Shi, A. Perrig, L. van Doorn, and P. Khosla. Pioneer: Verifying integrity and guaranteeing execution of code on legacy platforms. In *ACM Symposium on Operating Systems Principles (SOSP)*, Oct. 2005.

[76] S. Setty, R. McPherson, A. J. Blumberg, and M. Walfish. Making argument systems for outsourced computation practical (sometimes). In *Network and Distributed System Security Symposium (NDSS)*, Feb. 2012.

[77] D. Shasha and M. Snir. Efficient and correct execution of parallel programs that share memory. *ACM Transactions on Programming Languages and Systems (TOPLAS)*, 10(2):282–312, Apr. 1988.

[78] S. Shinde, D. Le Tien, S. Tople, and P. Saxena. Panoply: Low-tcb linux applications with sgx enclaves. In *Network and Distributed System Security Symposium (NDSS)*, Feb. 2017.

[79] R. Sinha, M. Costa, A. Lal, N. P. Lopes, S. Rajamani, S. A. Seshia, and K. Vaswani. A design and verification methodology for secure isolated regions. June 2016.

[80] E. G. Sirer, W. de Bruijn, P. Reynolds, A. Shieh, K. Walsh, D. Williams, and F. B. Schneider. Logical attestation: An authorization architecture for trustworthy computing. In *ACM Symposium on Operating Systems Principles (SOSP)*, Oct. 2011.

[81] R. T. Snodgrass and I. Ahn. Temporal databases. *IEEE Computer*, 19(9):35–42, Sept. 1986.

[82] Y. Tang and J. Yang. Secure deduplication of general computations. In *USENIX Annual Technical Conference*, July 2015.

[83] H.-W. Tseng and D. M. Tullsen. Data-triggered threads: Eliminating redundant computation. In *IEEE International Symposium on High Performance Computer Architecture (HPCA)*, Feb. 2011.

[84] J. Tucek, W. Xiong, and Y. Zhou. Efficient online validation with delta execution. In *ACM International Conference on Architectural Support for Programming Languages and Operating Systems (ASPLOS)*, Mar. 2009.

[85] G. Urdaneta, G. Pierre, and M. Van Steen. Wikipedia workload analysis for decentralized hosting. *Computer Networks*, 53(11):1830–1845, 2009.

[86] K. Veeraraghavan, D. Lee, B. Wester, J. Ouyang, P. M. Chen, J. Flinn, and S. Narayanasamy. DoublePlay: Parallelizing sequential logging and replay. *ACM Transactions on Computer Systems (TOCS)*, 30(1):3, 2012.

[87] K. Vikram, A. Prateek, and B. Livshits. Ripley: Automatically securing web 2.0 applications through replicated execution. In *ACM Conference on Computer and Communications Security (CCS)*, Nov. 2009.

[88] R. S. Wahby, Y. Ji, A. J. Blumberg, abhi shelat, J. Thaler, M. Walfish, and T. Wies. Full accounting for verifiable outsourcing. In *ACM Conference on Computer and Communications Security (CCS)*, Oct. 2017.

[89] M. Walfish and A. J. Blumberg. Verifying computations without reexecuting them: from theoretical possibility to near practicality. *Communications of the ACM (CACM)*, 58(2):74–84, Feb. 2015.

[90] J. M. Wing and C. Gong. Testing and verifying concurrent objects. *Journal of Parallel and Distributed Computing*, 17:164–182, 1993.

[91] R. Wray. BT drops plan to use Phorm targeted ad service after outcry over privacy. *The Guardian*, July 2009. https://www.theguardian.com/business/2009/jul/06/btgroup-privacy-and-the-net.

[92] Z. Yang, M. Yang, L. Xu, H. Chen, and B. Zang. ORDER: Object centRic DEterministic Replay for Java. In *USENIX Annual Technical Conference*, June 2011.

[93] A. Yip, X. Wang, N. Zeldovich, and M. F. Kaashoek. Improving application security with data flow assertions. In *ACM Symposium on Operating Systems Principles (SOSP)*, Oct. 2009.

[94] C. Zamfir, G. Altekar, and I. Stoica. Automating the debugging of datacenter applications with ADDA. In *Dependable Systems and Networks (DSN)*, June 2013.

[95] F. Zhang, J. Chen, H. Chen, and B. Zang. Cloudvisor: retrofitting protection of virtual machines in multi-tenant cloud with nested virtualization. In *ACM Symposium on Operating Systems Principles (SOSP)*, Oct. 2011.

[96] Y. Zhang, D. Genkin, J. Katz, D. Papadopoulos, and C. Papamanthou. vSQL: Verifying arbitrary SQL queries over dynamic outsourced databases. In *IEEE Symposium on Security and Privacy*, May 2017.

# A  Proof of Correctness

This appendix states the definition of correctness and then proves that our audit algorithm (and associated report collection process) meets that definition. The algorithm is defined in Section 3 and extended with additional object types in Sections 4.4–4.5. For completeness, the algorithm is included in this appendix in Figure 12, where it is called SSCO_AUDIT2.

## A.1  Definition of correctness

The two correctness properties are Completeness and Soundness. We have described these properties (§2) and now make them more precise.

**Model.** We presume the setting of Section 2 and the concurrency model of Section 3.2. We use those definitions. Additionally, in this appendix, we will sometimes use the word "request" as a shorthand for "the execution of the program, on input given by the request"; there are examples of this use in the next several sentences.

Each request consists of a sequence of instructions. A special instruction allows a request to invoke an *operation* on a shared object, which we often call a *state item*. A request can execute any number of such state operations over the lifetime of request, but it can issue only one at a time. Our algorithm handles three kinds of state items (§3.2, §4.4): atomic registers [55]; linearizable key-value stores exposing single-key get/set operations; and strictly serializable databases, with the restrictions stated in Section 4.4.

It will be convenient to have a notation for events (requests and responses) in the trace. We represent such events as a tuple:

$$(\text{RESPONSE} \mid \text{REQUEST}, \textit{rid}, [\text{contents}])$$

A trace is a timestamped or ordered list of such tuples (the exact timing does not matter, only the relative order of events). We assume that traces are balanced: every response is associated to a request, and every request has exactly one response. In practice, the verifier can ensure this property prior to beginning the audit.

**Definition 1** (Completeness). *A report collection and audit procedure are defined to be* Complete *if the following holds. If the executor executes the program (under the model above) and the given report collection procedure, then the given audit procedure (applied to the resulting trace and reports) accepts.*

To define Soundness in our context requires several additional notions.

*Request schedule.* A request schedule models the thread context switch schedule, and is an ordered list of requestIDs. For example:

$$\text{req 1, req 23, req 1, req 14, req 5, req 1, \ldots}$$

Notice that requestIDs are permitted to repeat in the schedule.

*Operationwise execution.* Consider a (physically impossible) model where, instead of requests arriving and departing, the executor has access to all requestIDs in a trace and their inputs. Operationwise execution means executing the program against all requestIDs by following a request schedule. Specifically, for each request id (rid) in the request schedule, in order:

- If it is the rid's first appearance, the executor reads in that request's input and allocates the structures needed to run the program on that input.
- Otherwise, the executor runs the request up to and including the request's next interaction with the outside world. This will either be an operation on a state object, or the delivered output (as the response).

After each such event, the request is held, until the executor reschedules it. If a request is scheduled after it has delivered its response, the executor immediately yields and chooses the next rid in the request schedule.

*Request precedence.* A trace *Tr* induces a partial order on requests. We say that a request $r_1$ precedes another request $r_2$ in a trace if the trace shows that (the execution of) $r_1$ must have ended before (the execution of) $r_2$ began. We notate this relation as $<_{Tr}$. That is, $r_1 <_{Tr} r_2$ if the event (RESPONSE, $r_1$, ·) occurs in *Tr* before (REQUEST, $r_2$, ·).

*Real-time consistency.* A request schedule *S* of the kind specified above is *real-time consistent* with $<_{Tr}$ if for any $r_1, r_2$ that appear in *S*, $r_1 <_{Tr} r_2 \implies$ all instances of $r_1$ in *S* are sequenced before all instances of $r_2$ in *S*.

Now we can define Soundness:

**Definition 2** (Soundness). *A report collection and audit procedure are defined to be* Sound *if the following holds. If the given audit procedure accepts a trace Tr and reports R, then there exists some request schedule, S, such that:*

*(a) The outputs produced by operationwise execution according to S (on the inputs in Tr) are exactly the responses in the trace Tr, and*

*(b) S is real-time consistent with $<_{Tr}$.*

**Comments on the Soundness definition.** The Soundness definition is bulky, and it may be unintuitive. But it is capturing something natural: it says that a trace *Tr* and reports can pass the given audit process only if the *Tr* is consistent with having actually executed the requests in *Tr*, according to a physical schedule at the executor.

Here is a potentially more intuitive description. The verifier accepts only if the following holds: there exists a schedule *S* such that if we feed requests to a well-behaved executor, in the exact order that they arrived, and if the executor physically context-switches among them according to *S*, then the executor will emit the precise responses that are in the trace, in the precise order (relative to the requests and the other responses) that they appear in the trace.

One might wonder why do we not simply phrase the definition as, "If the audit procedure accepts, then it means that the executor executed the program." The reason is that making this determination in our model is impossible: the trace collector, and the verifier, have no visibility to look "inside" the executor. For all they can tell, the executor is passing their checks by doing something radically different from executing the program. The key point, however, is that if the verifier accepts, it means that what the executor actually did and what it is supposed to be doing are completely indistinguishable from outside the executor.

The definition does contain some leeway for an untrusted executor: any physical schedule that it chooses will result in the verifier accepting (provided the executor actually executes the requests according to that schedule). But that leeway seems unavoidable: in this execution and concurrency model, even a well-behaved executor has complete discretion over the schedule.

**Model vs. proofs.** Our pseudocode and proofs assume the model stated at the outset of this section, but with one simplification. Specifically, the pseudocode and the proofs treat a database transaction as a series of statements, with no code interspersed (see, for example, line 27 in Figure 12). We impose the simplification to avoid some tedium in the proofs. However, the model itself reflects the implementation (§4), which does permit code (for example, PHP) to execute in between SQL statements that are part of the same transaction. We address the complexity in Section A.7.

**Model vs. reality.** A major difference between our model and web applications is that we are assuming that state operations and scheduling are the sole sources of non-determinism. In reality, there is another source of non-determinism: the return values of some functions (those that query the time, etc.). To take this into account would complicate the definitions and proofs; for example, Soundness and Operationwise execution would need to explicitly refer to the reports that hold the return values of non-deterministic functions (§4.6). To avoid this complexity, our claims here ignore this source of non-determinism.

### A.2 Proof outline and preliminaries

Our objective is to prove that SSCO_AUDIT2 (Figure 12), together with the corresponding report collection process (§3–§4), meets Completeness (Definition 1) and Soundness (Definition 2). Below, we outline the main idea of the proof. However, this is just a rough description; the proofs themselves go in a different order (as befits the technique).

- A central element in the proofs is a variant of SSCO_AUDIT2 that we define, called OOOAudit (§A.4). This variant relies on out-of-order, simulated execution, which we call OOOExec (in contrast to ReExec2, Figure 12). OOOExec follows some supplied op schedule $S$ (a list of requestIDs), and executes one request at a time up through the request's next op (rather than the grouped execution; §3.1). $S$ is required to respect program order. Thus, while requests can be arbitrarily interleaved in $S$, each executes sequentially.

- We establish that OOOExec (following some schedule $S$) is equivalent to ReExec2 (following some control flow reports $C$); the argument for this is fairly natural and happens at the end (§A.6).

The proof focuses on OOOAudit. The rough argument for the Soundness of OOOAudit is as follows (§A.5). Assume that OOOAudit accepts on some schedule $S$.

- We establish that OOOAudit is indifferent to the schedule (§A.4): following $S_1$ is equivalent to following any other schedule $S_2$, provided both respect program order.

- Let schedule $S'$ be an ordering (a topological sort) of the graph $G$. An ordering exists because $G$ has no cycles (otherwise, OOOAudit would not have accepted $S$). We establish that $G$, and hence $S'$, respects the partial order given by externally observable events (§A.3). By the previous bullet, OOOAudit accepts when following $S'$.

- We establish that this simulated execution (that is, OOOAudit following $S'$) is equivalent to physical execution according to $S'$. The idea underlying the argument is that the checks in the simulated execution (which pass) ensure that the op parameters in the logs, and the ops in the graph, match physical execution. That in turn means that simulating certain operations (such as reads) by consulting the logs produces the same result as in the physical execution.

- Meanwhile, the final check of OOOAudit (which, again, passes) establishes that the produced outputs (produced by the simulated execution) equal the responses in the trace. This bullet and the previous together imply that physical execution according to $S'$ produces the responses in the trace. Combining with the second bullet, we get that $S'$ is a schedule of the kind required by the soundness definition.

The argument for Completeness of OOOAudit uses similar reasoning to the argument for Soundness (§A.5). The essence of the argument is as follows. If the executor operates correctly, then (1) the reports supplied to the verifier constitute a precise record of online execution, and (2) OOOAudit, when supplied $S'$, reproduces that same online execution. As a result, the contents that OOOAudit expects to see in the reports are in fact the actual contents of the reports (both reflect online execution). Therefore, the checks in OOOAudit pass. That implies that OOOAudit accepts any schedule that respects program order.

| **Input** Trace *Tr* | **Input** Reports *R* | **Global** *OpMap*: (requestID, opnum) → (*i*, seqnum) | **Global** *kv*, *db*   // versioned storage |

The form of the opcontents depends on the optype:

| optype | opcontents | optype | opcontents |
|---|---|---|---|
| RegisterRead | empty | KvSet | key and value to write |
| RegisterWrite | value to write | DBOp | SQL statement(s), whether succeeds |
| KvGet | key to read | | |

Components of the reports *R*:
*C*: CtlFlowTag → Set(requestIDs)  // purported groups; §3.1
$OL_i$: $\mathbb{N}^+$ → (requestID, opnum, optype, opcontents)  // §3.3
*M*: requestID → $\mathbb{N}$   // op counts; §3.3

```
 1: procedure SSCO_AUDIT2()
 2:    // Partially validate reports (§3.5) and construct OpMap
 3:    ProcessOpReports()      // defined in Figure 5
 4:
 5:    kv.Build(OL_{i_kv})     // OL_{i_kv} is op log for versioned key-value store (§4.5)
 6:    db.Build(OL_{i_db})     // OL_{i_db} is op log for versioned database (§4.5)
 7:
 8:    return ReExec2 ()       // line 29
 9:
10: procedure CHECKOP(rid, opnum, i, optype, opcontents)
11:    if (rid, opnum) not in OpMap: REJECT
12:    î, s ← OpMap[rid, opnum]
13:    ôt, ôc ← (OL_i[s].optype, OL_i[s].opcontents)
14:    if i ≠ î or optype ≠ ôt or opcontents ≠ ôc: REJECT
15:    return s
16:
17: procedure SIMOP(i, s, optype, opcontents)
18:    ret ← ⊥
19:    if optype = RegisterRead:
20:       writeop ← walk backward in OL_i from s; stop when
21:          optype=RegisterWrite
22:       if writeop doesn't exist: REJECT
23:       ret = writeop.opcontents
24:    else if optype = KvGet:
25:       ret = kv.get(opcontents.key, s)
26:    else if optype = DBOp:
27:       ret = db.do_trans(opcontents.transaction, s)
28:    return ret

29: procedure REEXEC2()
30:    Re-execute Tr in groups according to C:
31:
32:    (1) Initialize a group as follows:
33:        Read in inputs for all requests in the group
34:        Allocate program structures for each request in the group
35:        opnum ← 1    // opnum is a per-group running counter
36:
37:    (2) During SIMD-on-demand execution (§3.1):
38:
39:        if execution within the group diverges: return REJECT
40:
41:        When the group makes a state operation:
42:            optype ← the type of state operation
43:            for all rid in the group:
44:                i, oc ← state op parameters from execution
45:                s ← CheckOp(rid, opnum, i, optype, oc)  // line 10
46:                if optype ∈ {RegisterRead, KvGet, DBOp}:
47:                    state op result ← SimOp(i, s, optype, oc)  // line 17
48:            opnum ← opnum + 1
49:
50:    (3) When a request rid finishes:
51:        if opnum < M(rid): return REJECT
52:
53:    (4) Write out the produced outputs
54:
55:    if the produced outputs from (4) are exactly the responses in Tr:
56:        return ACCEPT
57:    return REJECT
```

**Figure 12**—SSCO audit procedure. This is a refinement of Figure 3. This one includes additional objects: a versioned database and key-value store, as used by OROCHI (§4.5). As in Figure 3, the Trace *Tr* is assumed to be balanced.

**Conventions and notation.** Per Section 3 and Figure 12, the reports *R* have components $C, M, OL_1, \ldots, OL_n$, where *n* is the number of state items. For convenience, we will use this notation, rather than *R.M*, *R.OL_i*, etc. Note that for a DB log, the opcontents is the SQL statement(s) in a transaction. For example, if a DB is labeled with $i = 3$, then $OL_3[j]$.opcontents contains the SQL statements in the $j^{\text{th}}$ DB transaction.

A ubiquitous element in the proofs is the graph *G* that is constructed by ProcessOpReports. *G* depends on *Tr* and *R*, but we will not notate this dependence explicitly. Likewise, when notating directed paths in *G*, we leave *G* implicit; specifically, the notation $p \rightsquigarrow q$ means that there is directed path from node *p* to node *q* in graph *G*.

### A.3  Ordering requests and operations

Our first lemma performs a bit of advance housekeeping; it relates the graph *G*, the Op Count reports (*M*), and the operation log reports (*OL*). (The lemma is phrased in terms of *OpMap*, rather than operation log reports, but in a successful run, *OpMap* contains one entry for each log entry, per Figure 5, line 38.) The lemma says that, if ProcessOpReports succeeds, then *G* "matches" *OpMap*, in the sense that every operation label in *G* has an entry in *OpMap*, and every entry in *OpMap* appears in *G*. This in turn means that there is a correspondence between the nodes of *G* and the operations in the logs.

Note that if ProcessOpReports succeeds, it does not mean that *G* or the operation logs are "correct"; overall, they could be nonsense. For example, *M(rid)* could be too big or too small for a given *rid*, meaning that there are spurious or insufficient entries in the logs (and *G*). Or the operations of a given *rid* could be in the wrong operation log (meaning that the reports include a wrong claim about which state item a given operation targets). Or the logged contents of a write operation could be spurious. There are many other cases besides. All of them will be detected during re-execution. Importantly, we do not need to enumerate the forms of incorrectness. Rather, we let the proofs establish end-to-end correctness.

**Lemma 1** (Report consistency). *If ProcessOpReports accepts, then the domain of OpMap (which is all entries in*

*the log files) is exactly the set*

$$T = \{(rid, j) \mid rid \text{ is in the trace and } 1 \leq j \leq M(rid)\},$$

*and*

$$G.Nodes = T \cup \{(rid, 0), (rid, \infty) \mid rid \text{ is in the trace}\}.$$

*Proof.* Take $(rid, j)$ in $T$. By definition of $T$, $rid$ is in the trace and $1 \leq j \leq M(rid)$. The final check in CheckLogs (which passed, per the premise) considers this element, and ensures that it is indeed in *OpMap*. Now consider the domain of *OpMap*. An element $(rid, j)$ can be inserted into *OpMap* (Figure 5, line 38) only if $rid$ is in the trace, $j > 0$, and $j \leq M(rid)$ (lines 31–33).

The second part of the lemma is immediate from the logic in AddProgramEdges (Figure 5). □

The remaining lemmas in this section establish that the order in $G$ is consistent with externally observable events, as well as the claimed ordering in the operation logs. The first step is to prove that the graph $G_{Tr}$ produced by CreateTimePrecedenceGraph (Figure 6) explicitly materializes the $<_{Tr}$ relation. We use $r_1 \prec r_2$ to denote that there is a directed path from $r_1$ to $r_2$ in $G_{Tr}$.

**Lemma 2** (Correctness of CreateTimePrecedenceGraph). *For all $r_1, r_2$ in $Tr$, $r_1 <_{Tr} r_2 \iff r_1 \prec r_2$.*

*Proof.* We begin with the $\implies$ direction. Take $r_1, r_2$ with $r_1 <_{Tr} r_2$. Consider a sequence $T$

$$r_1 <_{Tr} s_1 <_{Tr} \cdots <_{Tr} s_n <_{Tr} r_2$$

that is "tight" in that one cannot insert further elements that obey $<_{Tr}$ between the members of this sequence. (At least one such sequence must exist.) We claim that there is a directed path:

$$r_1 \prec s_1 \prec \cdots \prec s_n \prec r_2.$$

Now, if for all adjacent elements $t, u$ in sequence $T$, there is an edge $\langle t, u \rangle$ in $G_{Tr}$, then the claim holds.

Assume toward a contradiction that there are adjacent $t, u$ without such an edge. Then, at the time that CreateTimePrecedenceGraph (Figure 6) processes the event (REQUEST, $u$, ·), request $t$ must have been already evicted from Frontier (if $t$ had not been evicted, then line 10 would have created the edge $\langle t, u \rangle$). This eviction must have been caused by some request $v$. But this implies that (RESPONSE, $t$, ·) precedes (REQUEST, $v$, ·) in the trace.[4] Furthermore, (RESPONSE, $v$, ·) precedes (REQUEST, $u$, ·) (because the eviction occurred before CreateTimePrecedenceGraph handled (REQUEST, $u$, ·)). Summarizing, there is a request $v$ for which:

$$t <_{Tr} v <_{Tr} u,$$

which contradicts the assumption that $t$ and $u$ were adjacent in sequence $T$.

---
[4]The detailed justification is that the eviction could have happened only if there is an edge $\langle t, v \rangle$ (by line 13); such an edge can exist only if $t$ was in the Frontier when CreateTimePrecedenceGraph processed (REQUEST, $v$, ·) (by line 10); and $t$ entered the Frontier after CreateTimePrecedenceGraph processed (RESPONSE, $t$, ·) (by line 14).

For the $\impliedby$ direction, consider $r_1, r_2$ with $r_1 \prec r_2$. If $r_1 \not<_{Tr} r_2$, then the directed path includes an edge $e = \langle s_1, s_2 \rangle$ for which $s_1 \not<_{Tr} s_2$; this follows from the fact that $<_{Tr}$ is transitive. Now, consider the point in CreateTimePrecedenceGraph at which $e$ was added (line 10). At that point, $s_1$ was in Frontier, which implies that (RESPONSE, $s_1$, ·) was observed already in the scan. This implies that (RESPONSE, $s_1$, ·) precedes (REQUEST, $s_2$, ·) in the trace, which means $s_1 <_{Tr} s_2$: contradiction. □

**Lemma 3** (G obeys request precedence). *At the end of ProcessOpReports, $r_1 <_{Tr} r_2 \iff (r_1, \infty) \rightsquigarrow (r_2, 0)$.*

*Proof.* $\implies$ : This follows from the proof of the prior lemma, the application of SplitNodes (Figure 5) to $G_{Tr}$, and the fact that for each $s_i$ in $T$, $(s_i, 0) \rightsquigarrow (s_i, \infty)$ (which itself follows from the program edges, added in Figure 5, lines 25–26).

$\impliedby$ : Consider $r_1, r_2$ with $(r_1, \infty) \rightsquigarrow (r_2, 0)$. If $r_1 \not<_{Tr} r_2$, then the directed path includes an edge $e = \langle (s_1, \infty), (s_2, 0) \rangle$ for which $s_1 \not<_{Tr} s_2$; this follows from the fact that $<_{Tr}$ is transitive. But if $e$ is an edge in $G$, then $\langle s_1, s_2 \rangle$ is an edge in $G_{Tr}$, which implies, by application of Lemma 2, that $s_1 <_{Tr} s_2$: contradiction. □

**Lemma 4** (G obeys log precedence). *If ProcessOpReports accepts, then for all operation logs $OL_i$,*

$$1 \leq j < k \leq \text{length}(OL_i) \implies$$
$$(OL_i[j].\text{rid}, OL_i[j].\text{opnum}) \rightsquigarrow (OL_i[k].\text{rid}, OL_i[k].\text{opnum}).$$

*Proof.* Fix $OL_i, j$; induct over $k$. Base case: $k = j + 1$. If $OL_i[j].\text{rid} = OL_i[j+1].\text{rid}$, then the check in Figure 5, line 54 and the existing program edges together ensure that $(OL_i[j].\text{rid}, OL_i[j].\text{opnum}) \rightsquigarrow (OL_i[j+1].\text{rid}, OL_i[j+1].\text{opnum})$. If on the other hand $OL_i[j].\text{rid} \neq OL_i[j+1].\text{rid}$, then line 53 in AddStateEdges inserts an edge between the two nodes.

Inductive step: consider $k + 1$. Reasoning identical to the base case gives $(OL_i[k].\text{rid}, OL_i[k].\text{opnum}) \rightsquigarrow (OL_i[k+1].\text{rid}, OL_i[k+1].\text{opnum})$. And the induction hypothesis gives $(OL_i[j].\text{rid}, OL_i[j].\text{opnum}) \rightsquigarrow (OL_i[k].\text{rid}, OL_i[k].\text{opnum})$. Combining the two paths establishes the result. □

```
 1: Global Trace Tr, Reports R      // includes OL_i
 2:
 3: // S is an op schedule (§A.4)
 4: procedure OOOEXEC(S)
 5:     for each (rid, opnum) in S:
 6:         if opnum = 0:
 7:             Read in inputs from request rid in Tr
 8:             Allocate program structures for a thread to run rid
 9:
10:         else if opnum = ∞: // check that the thread produces output
11:             Run rid's allocated thread until the next event.
12:             If the event is a state operation or silent exit:
13:                 return REJECT
14:             Write out the produced output
15:
16:         else
17:             Run rid up to, but not including the next event; if the
18:             event is not a state operation, return REJECT
19:
20:             i, optype, oc ← state op parameters from execution
21:             s ← CheckOp(rid, opnum, i, optype, oc)
22:             if optype ∈ {RegisterRead, KvGet, DBOp}:
23:                 state op result ← SimOp(i, s, optype, oc)
24:
25:     if all produced outputs exactly match the responses in Tr:
26:         return ACCEPT
27:     return REJECT
```

**Figure 13**—Definition of OOOExec, a variant of ReExec2 (Figure 12) that executes according to an op schedule (§A.4). A central concept in the correctness proofs is OOOAudit, which is the same as SSCO_AUDIT2 (Figure 12), except that ReExec2 is replaced with OOOExec.

### A.4  Op schedules and OOOAudit

A lot of the analysis in the proof is with respect to a hypothetical audit procedure, which we call OOOAudit, that is a variant of SSCO_AUDIT2. OOOAudit performs out-of-order execution of requests but not in a grouped way (as in §3.1); the specifics are given shortly. OOOAudit relies on an augmented notion of a request schedule, defined below, in which requests are annotated with a per-request op number (or infinity). These annotations are not algorithmically necessary; they are a convenience for the proofs.

**Definition 3** (Op schedule). *An op schedule is a map:*

$$S: \mathbb{N} \to RequestId \times (\mathbb{N} \cup \{\infty\}).$$

*For example,*

$$(1,0), (23,0), (1,1), (23,1), (23,2), (23,\infty), (1,2), \ldots$$

**Definition 4** (Well-formed op schedule). *An op schedule S is* well-formed *(with respect to a trace Tr and set of reports R) if (a) S is a permutation of the nodes of the graph G that is constructed by ProcessOpReports, and (b) S respects program order.*

**Definition 5** (OOOAudit). *Define a procedure* OOOAudit*(Trace Tr, Reports R, OpSched S) that is the same as* SSCO_AUDIT2*(Trace Tr, Reports R), except that*

$$ReExec2() \quad (\text{Figure 12, line 29})$$

*is replaced with*

$$OOOExec(S) \quad (\text{Figure 13})$$

**Lemma 5** (Equivalence of well-formed schedules). *For all op schedules $S_1, S_2$ that are well-formed (with respect to Tr and R),*

$$OOOAudit(Tr, R, S_1) = OOOAudit(Tr, R, S_2).$$

*Proof.* Consider both invocations, one with $S_1$ and one with $S_2$. The schedule ($S_1$ versus $S_2$) does not affect OOOAudit until the line that invokes OOOExec. So ProcessOpReports fails in both executions, or neither. Assume that these procedures succeed, so that both executions reach OOOExec. We need to show that $OOOExec(S_1) = OOOExec(S_2)$.

$S_1$ and $S_2$ have the same operations, because they are both constructed from the same graph G. Meanwhile, OOOExec re-executes each request (meaning each rid) in isolation. To see this, notice that none of the lines of OOOExec modifies state that is shared across requests (*OpMap*, *kv*, etc.). Therefore, the program state (contents of PHP variables, current instruction, etc.) of a re-executed request rid evolves deterministically from operation to operation, and hence the handling of each operation for each rid is deterministic. This holds regardless of where the operations of an rid appear in an op schedule, or how the operations are interleaved.

Now, if $OOOExec(S_1)$ accepts, then all checks pass, and all produced outputs match the responses in the trace. The preceding paragraph implies that $OOOExec(S_2)$ would encounter the same checks (and pass them), and produce the same outputs. On the other hand, if $OOOExec(S_1)$ rejects, then there was a discrepancy in one of the checks or in the produced output. $OOOExec(S_2)$ either observes the same discrepancy, or else it rejected earlier than this, where "early" is with reference to the sequencing in $S_2$. Summarizing, $OOOExec(S_1)$ and $OOOExec(S_2)$ deliver the same accept or reject decision, as was to be shown. □

### A.5  Soundness and completeness of OOOAudit

**Lemma 6** (OOOAudit Soundness). *If there exists a well-formed op schedule S for which OOOAudit(Tr, R, S) accepts, then there exists a request schedule $S''$ with properties (a) and (b) from Definition 2 (Soundness).*

*Proof.* If OOOAudit(Tr, R, S) accepts, then there are no cycles in the graph G produced by ProcessOpReports (OOOAudit calls into ProcessOpReports, and ProcessOpReports—specifically lines 11–12 in Figure 5—would reject if there were a cycle). This means that G can be sorted topologically. Let the op schedule $S'$ be such a topological sort. Define the request schedule $S''$ to be the same as $S'$ but with the opnum component discarded.

$S'$ is well-formed: it contains the operations of $G$, and it respects program order because there are edges of $G$ between every two state operations in the same request. By Lemma 5, OOOAudit($Tr, R, S'$) returns accept.

*Property (b).* Observe that no (*rid*, *opnum*) appears twice in $S'$; this follows from the construction of $S'$ and $G$. Thus, one can label each (*rid*, *opnum*) in $S'$ with its sequence number in $S'$; call that labeling *Seq*. Also, note that for nodes $n_1, n_2$ in $G$, if $n_1 \rightsquigarrow n_2$, then $Seq(n_1) < Seq(n_2)$; this is immediate from the construction of $S'$ as a topological sort of $G$, and below we refer to this fact as "$\rightsquigarrow$ implies $<$".

Now, assume to the contrary that $S''$ does not meet property (b) in Definition 2. Then there exist $r_1, r_2$ with $r_1 <_{Tr} r_2$ and at least one appearance of $r_2$ occurring in $S''$ before at least one appearance of $r_1$. In that case, $S'$ must contain $(r_1, i), (r_2, j)$ such that $Seq(r_2, j) < Seq(r_1, i)$. Thus, we have:

$$\begin{aligned}
Seq(r_2, j) &< Seq(r_1, i) &\text{[from contrary hypothesis]} \\
&\leq Seq(r_1, \infty) &\text{[}\rightsquigarrow \text{ implies } <\text{]} \\
&< Seq(r_2, 0) &\text{[Lemma 3; } \rightsquigarrow \text{ implies } <\text{]} \\
&\leq Seq(r_2, j) &\text{[}\rightsquigarrow \text{ implies } <\text{]}
\end{aligned}$$

which is a contradiction.

*Property (a).* We establish this property by arguing, first, that re-executing (according to the op schedule $S'$) is the same as a physical (online) execution, in which the request scheduling is given by $S'$. This is the longest (and most tedious) step in the proof of soundness. Second, we argue that such a physical execution is equivalent to the earlier notion of operationwise execution (§A.1). To make these arguments, we define two variants of the audit immediately below, and then prove the two equivalences in sub-lemmas:

*Actual.* Define a variant of OOOAudit($Tr, R, S$) called *Actual*(Trace $Tr$, Reports $R$, OpSched $S$). In Actual, there is a physical state object $i$ for each operation log $OL_i$. Execution in Actual proceeds identically to execution in OOOAudit, except that state operations are concretely executed, instead of simulated. Specifically, Actual invokes CheckOp but then, instead of simulating certain operations (Figure 13, lines 22–23), it performs all operations against the corresponding physical state object.

*Operationwise.* Define *Operationwise*(Trace $Tr$, RequestSched $RS$) to be the same as Actual, except that:

(i) All checks, including CheckOp, are discarded (notice that the signature of Operationwise does not include the reports that would enable checks).

(ii) Operationwise is not presented with opnums (notice that the schedule argument $RS$ is a request schedule, not an op schedule). Instead, Operationwise simulates opnums:

when an rid first appears in $RS$, Operationwise does what Actual does when the opnum is 0 (it reads in the inputs, allocates program structures, etc.). Subsequent appearances of rid cause execution through the next operation or the ultimate output.

**Sub-lemma 6a.** *OOOAudit($Tr, R, S'$) and Actual($Tr, R, S'$) produce the same outputs.*

*Proof.* We will argue that every schedule step preserves program state in the two runs. Specifically, we claim that for each (*rid*, *opnum*), both runs have the same state at line 24 (Figure 13). We induct over the state operations in $S'$, turning to the inductive step first.

*Inductive step.* Consider a state operation in $S'$; it has the form $(rid, j)$, where $j \in \{1, 2, \ldots, M(rid)\}$. The induction hypothesis is that for all entries before $(rid, j)$ in $S'$, the two runs have the same state in line 24 (Figure 13).

If $j = 1$, note that execution proceeds deterministically from thread creation, so lines 20–22 execute the same in the two runs. If $j > 1$, then execution of operation $(rid, j-1)$ was earlier in $S'$ and execution of *rid* "left off" at line 24. The induction hypothesis implies that, at that point, the state in the two runs was the same. From that point, through the current operation's lines 20–22, execution is deterministic. In both cases ($j = 1$ and $j > 1$), CheckOp (line 21) passes in Actual; this is because it passes in OOOAudit (which we know because, as established at the beginning of the proof of the overall lemma, OOOAudit($Tr, R, S'$) accepts), and Actual and OOOAudit invoke CheckOp with the same parameters.

It remains to show that, if *optype* $\in$ {RegisterRead, KvGet, DBOp}, then both runs read the same value in line 23. To this end, we establish below a correspondence between the history of operations in Actual and OOOAudit. Let $\hat{i}, \hat{s} \leftarrow OpMap[rid, j]$. Because CheckOp passes in both Actual and OOOAudit (with the same parameters in both runs), $i = \hat{i}$, and thus both Actual and OOOAudit will interact with the same logical object (Actual does so via physical operations; OOOAudit consults the corresponding log). We refer to this object as $i$ below.

*Claim.* Define $Q$ as $(OL_i[1], \ldots, OL_i[\hat{s}-1])$; if $\hat{s} = 1$, $Q$ is defined to be empty. Then $Q$ describes the operations, in order, that Actual issues against physical state object $i$, prior to $(rid, j)$.

*Proof.* We will move back and forth between referring to operations by their location in a log ($OL_i[k]$) and by (*rid*, *opnum*) (the domain of *OpMap*). There is no ambiguity because CheckLogs (Figure 5) ensures a bijection between log entries and (*rid*, *opnum*) pairs.

Each of the elements of $Q$, meaning each ($OL_i[k]$.rid, $OL_i[k]$.opnum), $1 \leq k \leq \hat{s}-1$, appears in $S'$ before the current operation (in an order that respects the log sequence numbers $1, \ldots, \hat{s}-1$); this follows from Lemma 4 (and the fact that $S'$ is a topological sort). Furthermore, in OOOAudit, these are the *only* operations in $S'$ (before the current operation) that interact with $OL_i$. To establish this, assume to the contrary that there is an additional operation $(rid', j')$ that appears in $S'$ before the current operation, with $OpMap[rid', j'] = (i, t)$, for some $t$. If $t \leq \hat{s}$, that violates the aforementioned bijection; if $t > \hat{s}$, that violates Lemma 4.

Now, consider the execution in Actual. If the history of operations to the corresponding physical state object does not match $Q$, then there is an operation in the relevant prefix of $S'$ for which the two runs diverge. Consider the *earliest* such divergence (that is, Actual and OOOAudit are tracking each other up to this operation); call that earliest diverging operation $(rid^*, j^*)$.

Consider what happens when $(rid^*, j^*)$ executes. Both runs produce $i^*, optype^*, oc^*$, the state op parameters yielded by execution (Figure 13, line 20). These three parameters are the same across Actual and OOOExec, by application of the induction hypothesis (again using reasoning as we did above: for operation $(rid^*, j^* - 1)$, program state was the same in line 24, etc.). Now, consider CheckOp (line 21). Both runs obtain $i', s' \leftarrow OpMap[rid^*, j^*]$, and $ot', oc' \leftarrow OL_{i^*}[s'].optype, OL_{i^*}[s'].opcontents$ (Figure 12, lines 12–13). But CheckOp passes in OOOExec (as argued earlier), and hence, $i^* = i', optype^* = ot', oc^* = oc'$. This means that the state operation issued by Actual corresponds precisely to what the log dictates (same logical object, same operation type, same parameters, etc.).

Thus, if the two runs diverge, it must be in the sequence number. In OOOExec, $(rid^*, j^*)$ causes operation $s'$ (ordinally) to log $OL_{i'}$. If the operation in Actual would be operation number $s^*$ (ordinally) to object $i'$, where $s^* \neq s'$, then there was an earlier divergence—either Actual did not issue an operation to object $i'$ when OOOExec issued an operation to $OL_{i'}$, or Actual issued an operation to object $i'$ when OOOExec did not issue an operation to $OL_{i'}$. But $(rid^*, j^*)$ was the earliest divergence, so we have a contradiction.  □

Now we must establish that the operations actually return the same values in Actual and OOOExec (in Figure 13, line 23). We begin with RegisterRead. For such operations, Actual returns the current value of the register; by register semantics, this value is that of the most recent "write" in time. Because $Q$ precisely reflects the history of operations in Actual (per the Claim), this most recent write is the RegisterWrite operation in $Q$ with the highest log sequence number. The contents of this operation is precisely what OOOAudit "reads" into program variables in SimOp (see Figure 12, line 21). Thus, OOOAudit and Actual read the same value into program variables (Figure 13, line 23).

Now let us consider what happens if *optype* is KvGet or DBOp. OOOAudit invokes either $kv.get(key, s)$ or $db.\text{do\_trans}(transaction, s)$ (Figure 12, lines 25 and 27). Each of these calls is equivalent to:

- Constructing state by replaying in sequence $OL_i[1], \ldots, OL_i[\hat{s} - 1]$ (specifically, the opcontents field of these log entries), and then
- Issuing the operation given by $OL_i[\hat{s}].opcontents$.

This equivalence is intuitively what *db* and *kv* provide, and we impose this equivalence as the required specification (see §A.7 for implementation considerations). Meanwhile, by the earlier Claim, the history of operations to object $i$ in Actual before the current operation is $OL_i[1], \ldots, OL_i[\hat{s}-1]$. Moreover, the current operation in Actual is given by *optype* and *oc* (Figure 13, line 20), which respectively equal $OL_i[\hat{s}].optype$ and $OL_i[\hat{s}].opcontents$; this follows from the fact that CheckOp passes for operation $(rid, j)$ in both executions. Therefore, Actual and OOOExec "see" equivalent histories and an equivalent current operation, for the state object in question. They therefore return the same result (Figure 13, line 23).

*Base case.* The first state operation in $S'$ has the form $(rid, 1)$. The reasoning proceeds identically to the inductive step, for $j = 1$. Here in the base case, $Q$ is always empty,[5] but this does not affect the logic.  □

**Sub-lemma 6b.** *Execution of program logic and state operations proceeds identically in* Actual$(Tr, R, S')$ *and* Operationwise$(Tr, S'')$. *In particular, they produce the same outputs.*

*Proof.* Actual$(Tr, R, S')$ passes all checks, so eliminating them does not affect the flow of execution. Furthermore, aside from the case opnum=0, the opnum component in $S'$ does not influence the flow of execution in Actual; the component only induces checks, which aren't run in Operationwise. For the opnum=0 case, notice that for each *rid*, $(rid, 0)$ always appears in $S'$ before $(rid, j)$, for any $j > 0$ or $j = \infty$ (this is because $S'$ is a topological sort of $G$). Thus, the treatment by Operationwise$(Tr, S'')$ of the first occurrence of *rid*—namely that it is as if Actual is encountering $(rid, 0)$—means that Operationwise and Actual execute this case identically.  □

To conclude, recall from the outset of the proof that OOOAudit$(Tr, R, S')$ accepts, which implies that it produces as outputs precisely the responses in the trace. Sub-lemmas 6a and 6b then imply that Operationwise$(Tr, S'')$ produces those outputs too. Meanwhile, Operationwise$(Tr, S'')$ has the precise form of operationwise execution (defined in Section A.1), which completes the argument.  □

**Lemma 7** (OOOAudit Completeness). *If the executor executes the given program (under the concurrency model given earlier) and the given report collection procedure, producing trace Tr and reports R, then for any well-formed op schedule S, OOOAudit$(Tr, R, S)$ accepts.*

---

[5]We know that $Q$ is empty, as follows. Let $\hat{i}, \hat{s} \leftarrow OpMap[rid, 1]$. If $\hat{s} > 1$, then operation $OL_{\hat{i}}[1]$ would have appeared earlier in $S'$, by Lemma 4. Therefore, $\hat{s} = 1$, which, by definition of $Q$, makes $Q$ empty.

*Proof.* Consider ProcessOpReports and OOOExec in turn.

**Sub-lemma 7a.** *ProcessOpReports passes.*

*Proof.* If the executor is well-behaved, then CheckLogs passes; this is because a well-behaved executor correctly sets $M$ and places each operation in exactly one log. Under those conditions, the checks in CheckLogs pass.

Now we need to show that CycleDetect (Figure 5, line 11) passes, i.e., we need to show that there are no cycles. If the executor is well-behaved, then there is a total ordering that defines when all log entries were written in the actual online execution; this is because entries are part of the "emissions" from a sequentially consistent execution. Furthermore, we can define in this total ordering "request begin" (which happens at the instant a request begins executing) and "request end" (which happens at the instant a request finishes executing). Notate these events as $(rid, 0)$ and $(rid, \infty)$, respectively. By sequential consistency, the $(rid, 0)$ event must precede all other $(rid, \cdot)$ events in the total ordering, and likewise $(rid, \infty)$ must succeed all other $(rid, \cdot)$ in the total ordering. Also, in the actual execution, if one request began after another ended, a well-behaved executor must have executed all operations for the former after all operations for the latter, so the total ordering respects that property too.

Now, in ProcessOpReports (Figure 5), an edge can be added to $G$ only in four cases (lines 53, 25, 26, and 10):

- An edge $(n_1, n_2)$ can be added to indicate that operation $n_1$ occurred before operation $n_2$, in the same log.
- An edge $(n_1, n_2)$ can be added to indicate that operation $n_1$ preceded operation $n_2$ in the same request.
- Edges for $\langle (rid, 0), (rid, 1) \rangle$ and $\langle (rid, m), (rid, \infty) \rangle$ are added, where $m$ is the purported maximum opnum for $rid$.
- If an edge of the form $\langle (r_1, \infty), (r_2, 0) \rangle$ is added, then $r_2$ began after $r_1$ ended.

Observe that in all four cases, an edge $\langle n_1, n_2 \rangle$ is added to the audit-time graph only if operation $n_1$ preceded operation $n_2$ in the total ordering given by the online execution. In other words, the graph edges are always consistent with the total ordering. Thus, if there is a cycle $n_1 \leadsto \cdots \leadsto n_1$ in $G$, it means that $n_1$ preceded itself in the total ordering, which contradicts the notion of total ordering. So there are no cycles. □

As established immediately above, $G$ has no cycles. It can therefore be topologically sorted.

**Sub-lemma 7b.** *Define op schedule $S'$ to be a topological sort of graph $G$. Then, the invocation $OOOExec(S')$:*
*(i) reproduces the program state of online execution, and*
*(ii) passes all checks*

*Proof.* Induct on the sequence $S'$.

Base case: The first operation in $S'$ has no ancestors in $G$. It is thus the first occurrence of its request and has the form $(rid, 0)$. $OOOExec(S')$ handles this by reading in input and allocating program structures deterministically; this is the same behavior as in the online execution.

Inductive step: assume that the claim holds for the first $\ell - 1$ operations in $S'$. Denote the op with sequence $\ell$ as $(rid, j)$. We reason by cases.

*Case I: $j = 0$.* Same reasoning as the base case.

*Case II: $j = \infty$.* Recall that $M$ is the Op Count report (§A.2). Consider the operation $(rid, M(rid))$; it appears in $S'$ prior to $(rid, j)$ and as the most recent entry for $rid$. This follows from the logic in ProcessOpReports and its callees. The induction hypothesis implies that program state in $OOOExec(S')$ is identical to the original online execution at $(rid, M(rid))$. This means that $OOOExec(S')$ will take the same next step that the original took, in terms of state operation versus exit versus output (because the original followed the program code, just as $OOOExec(S')$ is doing). Now, if that step were something other than an output, that would imply that $M(rid)$ was unfaithful to the online execution, contradicting the premise of a well-behaved executor. So the next interaction is indeed an output (in both executions), meaning that the check in $OOOExec(S')$ in the opnum=$\infty$ case passes (Figure 13, line 12). And the produced output is the same; in other words, the output produced by $OOOExec(S')$ is the same as what was produced online, which is what is in the trace. Thus, the output sameness check passes.

*Case III: $j = 1$.* By the induction hypothesis, $OOOExec(S')$ and the online execution had the same program state at $(rid, 0)$. This implies that $OOOExec(S')$ will take the same next step that the original took (in terms of state operation versus exit versus output). If that step is an output or exit rather than a state operation, that would imply that the executor inserted a spurious operation in the logs, contradicting the premise of a well-behaved executor. So the step is indeed an operation (in both executions). Being well-behaved, the executor recorded that operation as $(rid, 1)$ in the appropriate operation log, and this is the operation in question. Furthermore, the contents of the log entry (meaning the fields optype and opcontents) are faithful to the execution. Because of the determinism in passing from $(rid, 0)$ to $(rid, 1)$, the same program state is reproduced during $OOOExec(S')$, implying that all checks in CheckOp pass.

RegisterWrite and KvSet operations do not affect program state. Our remaining task under this case is to show that if the op has optype of RegisterRead, KvGet, or DBOp, then $OOOExec(S')$ produces the same value that the online execution did. To this end, let $(i, s) = OpMap[rid, 1]$, and consider the first $s-1$ operations to $OL_i$ in the original execution. These operations have been recorded as $OL_i[1], \ldots, OL_i[s-1]$, because the executor, being well-behaved, is tracking operations correctly. Thus, these log entries give the precise history to this state object (in the original execution) at the time of operation number $s$. (Note that this log could be any kind of log: register, key-value, etc.) Call the $s-1$ ops collectively $Q$. At this point, we can pick up the reasoning in the inductive step of Sub-lemma 6a after the Claim, only replacing "Actual" with "online execution".

*Case IV:* $1 < j \leq M(rid)$. $S'$ respects program order, so we can invoke the induction hypothesis on $(rid, j-1)$ to conclude that program state after executing $(rid, j-1)$ is the same in OOOExec($S'$) as it was when that operation was executed online. At this point, the same reasoning as in Case III applies, substituting $(rid, j)$ for $(rid, 1)$ and $(rid, j-1)$ for $(rid, 0)$. □

Sub-lemmas 7a and 7b imply that OOOAudit($Tr, R, S'$) accepts. Applying Lemma 5 to $S$ and $S'$ completes the proof. □

### A.6 Equivalence of OOOAudit and SSCO_AUDIT2

Having established the Completeness and Soundness of OOOAudit, it remains to connect the grouped executions (§3.1) to those of OOOAudit.

**Lemma 8.** *Given trace Tr and reports R, if* SSCO_AUDIT2($Tr, R$) *accepts, then there is a well-formed op schedule S that causes* OOOAudit($Tr, R, S$) *to accept.*

*Proof.* Recall that $C$ denotes the control flow grouping within $R$ (§A.2). One can construct $S$ as follows: Initialize $S$ to empty. Then run SSCO_AUDIT2($Tr, R$) (Figure 12), and every time SSCO_AUDIT2 begins auditing a control flow group $t$, add to $S$ entries $(rid, 0)$ for each $rid$ in the set $C(t)$. Whenever a group issues an operation (which, because the grouped execution does not diverge, the group does together), add $(rid, j)$ to $S$ for each $rid$ in $C(t)$, where $j$ is the running tally of opnums. When the requests write their output (which, again, they do together), add $(rid, \infty)$ to $S$ for each $rid$ in $C(t)$.

*Claim.* For each $rid$, the value of $j$ in $S$ prior to the $(rid, \infty)$ insertion is equal to $M(rid)$. Proof: SSCO_AUDIT2 accepted so passes the line that checks whether a request issues at least $M(rid)$ operations (Figure 12, line 51), implying that $j \geq M(rid)$. But $(rid, j)$ is in *OpMap* (otherwise line 11 would have rejected), and so, by Lemma 1, $j \leq M(rid)$. So this Claim is established. □

By the Claim, by the fact that $j$ increments (starting from 0), and by the fact that SSCO_AUDIT2's acceptance implies that all trace responses are produced (so all requests are executed), the constructed op schedule $S$ has all nodes from $G$. $S$ also respects program order. It is thus well-formed.

Meanwhile, executing OOOAudit($Tr, R, S$) would precisely replicate what happens in SSCO_AUDIT2($Tr, R$) because the only difference in execution is that the latter interleaves at the instruction and operand level, which does not affect program state; the flow and ordering is otherwise the same. This means that program state is also the same across the two algorithms at the time of issued operations, and hence the produced output is the same.

The checks are also the same. There is a superficial difference in how the "end state" is handled, but observe that both executions reject if a request $rid$ attempts to issue more than $M(rid)$ operations (in that case, the corresponding operation is not in *OpMap*, so CheckOp rejects, specifically line 11, Figure 12) or if a request attempts to exit, having issued fewer than $M(rid)$ operations (this happens in ReExec2 with an explicit check in Figure 12, line 51, and in OOOExec because if an operation produces output before the opnum=$\infty$ case, the algorithm rejects in Figure 13, lines 17–18).

Therefore, if all checks pass in SSCO_AUDIT2($Tr, R$), so do all checks in OOOAudit($Tr, R, S$), and OOOAudit($Tr, R, S$) accepts. □

Combining Lemma 8 with Lemma 6, we obtain SSCO_AUDIT2's soundness:

**Theorem 9** (SSCO_AUDIT2 soundness). *Given trace Tr and reports R, if* SSCO_AUDIT2($Tr, R$) *accepts, then there exists a request schedule with properties (a) and (b) from Definition 2 (Soundness).*

**Theorem 10** (SSCO_AUDIT2 completeness). *If the executor executes the given program (under the concurrency model given earlier) and the given report collection procedure, producing trace Tr and reports R, then* SSCO_AUDIT2($Tr, R$) *accepts.*

*Proof.* Use $C$ (the control flow grouping reports) to construct the following op schedule $S$: take each control flow group that SSCO_AUDIT2 would execute, and insert each request's operations in layers: first all of the opnum=0 entries appear for each $rid$ in the control flow group, then all of the opnum=1 entries, etc., up through $M(rid)$ for each $rid$ in the control flow group, and then all of the $(rid, \infty)$ entries, again for each $rid$ in the control flow group. Note that $M(rid)$ must be constant for all rids in a control flow group because otherwise $M$ is wrong or else the control flow grouping is not valid, either of which contradicts the executor being well-behaved.

$S$ respects program order, by construction. $S$ also includes all nodes from $G$. This follows because the executor is well-behaved, implying that $C$ includes all requestIDs in the trace. Meanwhile, $S$ includes $(rid, 0), (rid, 1), \ldots, (rid, M(rid)), (rid, \infty)$ for each of these rids, and those are exactly the nodes of $G$. Thus, $S$ is well-formed.

Lemma 7 implies that OOOAudit($Tr, R, S$) accepts. Compare the executions in SSCO_AUDIT2($Tr, R$) and OOOAudit($Tr, R, S$); the executions have the same logic, except for three differences:

(i) ReExec2 has an explicit check about whether a request issues fewer than $M(rid)$ operations (Figure 12, line 51),

whereas OOOExec has a separate opnum=∞ case (Figure 13, line 10).

(ii) ReExec2 executes a group from operation $j-1$ to operation $j$ in SIMD-style (§3.1) whereas OOOExec round-robins the execution from $j-1$ to $j$, for a group of requests.

(iii) ReExec2 rejects if execution diverges.

Difference (i) was handled in the proof of Lemma 8: the difference is superficial, in that both executions are requiring a request *rid* to issue exactly $M(rid)$ operations.

Difference (ii) does not result in different program state across the two executions. This is because any implementation of SIMD-on-demand (for example, OROCHI's acc-PHP; §4.3) is supposed to ensure that the SIMD-style execution is identical to executing each request in the group individually (as is done in OOOAudit($Tr, R, S$)), and so the results of all instructions (including op values, etc.) are the same between SSCO_AUDIT2($Tr, R$) and OOOAudit($Tr, R, S$).

For difference (iii), we have to argue that there is no divergence across requests within a control flow group in SSCO_AUDIT2($Tr, R$). Assume otherwise. Say that the divergence happens between ($rid, j$) and ($rid, j+1$), for one or more rids (in some control flow group), and consider the execution of all requests in the group up to ($rid, j$) in SSCO_AUDIT2($Tr, R$). The program state produced by this execution is equivalent to the program state at the same point in $S$ when executing OOOAudit($Tr, R, S$), because that is the whole point to the SIMD-style execution.

Consider now OOOAudit($Tr, R, S'$), where $S'$ is a topological sort of $G$ (we know that a topological sort exists because there are no cycles in $G$, and we know that there are no cycles in $G$ using the same reasoning as in Sub-lemma 7a). This execution results in the identical program state for each request as OOOAudit($Tr, R, S$), as argued in the proof of Lemma 5. But by Sub-lemma 7b, OOOAudit($Tr, R, S'$) reproduces the original online execution. This implies that if execution diverges during SSCO_AUDIT2($Tr, R$) for two requests in some control flow grouping, then the two requests had different executions during the original online execution. But if they did and if the executor placed them in the same control flow group, the executor is not well-behaved, in contradiction to the premise. □

### A.7 Details of versioned storage

**Key-value stores.** Recall the requirement referenced in the proof of Sub-lemma 6a: letting $i^*$ identify the key-value store object and its operation log, *invoking $kv.\text{get}(k, s)$ must be equivalent to creating a snapshot of a key-value store by replaying the operations $OL_{i^*}[1], \ldots, OL_{i^*}[s-1]$, and then invoking "get(k)" on that snapshot.*

To meet this requirement, OROCHI (§4) implements *kv* as a map from keys to (seq,value) pairs. The invocation $kv.\text{Build}(OL_{i^*})$ (Figure 12, line 5) constructs this map from all of the KvSet operations in $OL_{i^*}$. During re-execution, $kv.\text{get}(k, s)$ (Figure 12, line 25) performs a lookup on key $k$ to get a list of (seq,value) pairs, and then performs a search to identify, of these pairs, the one with the highest seq less than $s$ (or false if there is no such pair); $kv.\text{get}$ returns the corresponding value.

Summarizing, $kv.\text{get}(k, s)$ returns, of all of the entries in $OL_{i^*}$, the KvSet to key $k$ with highest sequence less than $s$. Meanwhile, if one were to replay $OL_i[1], \ldots, OL_i[s-1]$ to an abstract key-value store and then issue a "get(k)", one would get the most recent write—which is the same as the highest sequenced one in the set $OL_i[1], \ldots, OL_i[s-1]$. Thus, the implementation matches the requirement.

**Databases.** Transactions create some complexity. On the one hand, the pseudocode (Figures 12 and 13) and proofs treat multiple SQL statements in a transaction as if they are a single operation. On the other hand, in the implementation (and in the model given at the outset; §A.1), code can execute between the individual SQL statements of a transaction.

We briefly describe how to adapt the pseudocode and proofs to the actual system. Our point of reference will be Figure 12. As a bookkeeping detail, the system maintains a per-query unique *timestamp*. This identifier is not in the operation logs; it's constructed by the verifier. When building the versioned database (Figure 12, line 6), the verifier assigns each query the timestamp $ts = s \cdot \text{MAXQ} + q$, where $s$ is the enclosing transaction's sequence number in the operation log, MAXQ is the maximum queries allowed in one transaction (10000 in our implementation), and $q$ is the query number within the transaction. Another detail is that, for the database operation log, each entry's opcontents field is structured as an array of queries.

In the pseudocode, we alter lines 44–47 (in Figure 12). For DBOps, CheckOp and SimOp need to happen in a loop, interleaved with PHP execution. Instead of checking the entire transaction at once, these functions check the individual queries within the transaction. Specifically, using a query's timestamp, CheckOp and SimOp check whether each query produced by program execution is the same as the corresponding query in the operation log, and simulate the queries against versioned storage.

The proofs can regard program state as proceeding deterministically from query to query, in analogy with how the proofs currently regard program state proceeding deterministically from op to op. This is valid because, per the concurrency and atomic object model, there are no state operations interleaved with the enclosing transaction (§4.4, §A.1).

For the system and the proofs to make sense, the versioned database implementation has to meet the following requirement, which is analogous to that given for key-value stores earlier. Let $i^*$ identify the database object and its operation log.

- For timestamp *ts*, let $s = \lfloor ts/\text{MAXQ} \rfloor$ and let $q = ts \bmod \text{MAXQ}$; for convenience, let *queries* = $OL_i[s]$.opcontents.queries.
- The values returned by invoking *db*.do_query(*sql*, *ts*) must be equivalent to:
  — Creating a snapshot of a database by replaying the transactions $OL_{i^*}[1], \ldots, OL_{i^*}[s-1]$ followed by the queries *queries*$[1], \ldots,$ *queries*$[q-1]$, and then
  — Issuing the query *sql*.

To meet this requirement, OROCHI (§4) implements *db* atop a traditional SQL database, in a manner similar to WARP [27]. Specifically, the database used for the application is augmented with two columns: *start_ts* indicates when a given row was updated to its current value, and *end_ts* indicates when this row is updated to the next value. The invocation *db*.Build($OL_{i^*}$) (Figure 12, line 6) inserts rows with the relevant column values, using all of the queries in $OL_{i^*}$. During re-execution, *db*.do_query(*sql*, *ts*) obtains its results by passing *sql* to the underlying storage, augmented with the condition *start_ts* $\le ts <$ *end_ts*.

One can show that this implementation meets the requirement above, but the details are tedious.

### A.8 Efficiency of ProcessOpReports (time, space)

In this section, we analyze the time and space needed to execute ProcessOpReports (Figure 5) and the space needed to hold *OpMap*. Let $X$ be the total number of requests, $Y$ be the total number of state operations, and $Z$ be the cardinality of the minimum set of edges needed to represent the $<_{Tr}$ relation. Roughly speaking, the more concurrency there is, the higher $Z$ is. For intuition, if there are always $P$ concurrent requests, which arrive in $X/P$ epochs (so all requests in an epoch are concurrent with each other but succeed all of the requests in the prior epoch), then $Z \approx X \cdot P/2$ (every two adjacent epochs is a bipartite graph with all nodes on one side connecting to all nodes on the other).

**Lemma 11.** *The time and space complexity of ProcessOpReports are both $O(X + Y + Z)$. The space complexity of OpMap is $O(Y)$.*

*Proof.* We begin with time complexity. The graph $G$ is maintained as an adjacency list, so we assume that inserting a node or edge is a constant-time operation. ProcessOpReports first constructs *R.M*, at a cost of $O(X)$; this is not depicted.

After that, ProcessOpReports comprises six procedures: CreateTimePrecedenceGraph, SplitNodes, AddProgramEdges, CheckLogs, AddStateEdges, and CycleDetect.

To analyze CreateTimePrecedenceGraph, notice that, when handling a request's arrival, the algorithm iterates over *Frontier*, the number of iterations being equal to the number of edges connecting this edge to its predecessors. Similarly, when handling the request's arrival, the algorithm iterates over those same edges. So the total number of iterations has the same order complexity as the number of edges added; this is exactly $Z$, because CreateTimePrecedenceGraph adds the optimal number of edges (shown in the next claim). This implies that CreateTimePrecedenceGraph runs in time $O(X+Z)$.

SplitNodes performs a linear pass over the nodes and edges of $G_{Tr}$ so runs in time $O(X + Z)$.

AddProgramEdges and CheckLogs each perform at least one iteration for each state operation and each request, so these are both $O(X + Y)$. AddStateEdges iterates over every state operation in the logs, so it is $O(Y)$.

The dominant cost is CycleDetect. This is done with a standard depth-first search [32, Ch. 22], which is $O(V +E)$, where $V$ is the number of vertices and $E$ is the number of edges in the graph $G$. In our context, $V = 2 \cdot X + Y$, because each state op has a vertex, and we have the $(\cdot, 0)$ and $(\cdot, \infty)$ vertices for each *rid*. To upper-bound $E$, let us analyze each vertex type. The edges into $(rid, 0)$ and out of $(rid, \infty)$ are "split" from the original $Z$ edges that CreateTimePrecedenceGraph added to $G_{Tr}$; additionally, the out-edges from the $(rid, 0)$ vertices and the in-edges to the $(rid, \infty)$ vertices add an additional $2X$ edges total. An op vertex can have 4 edges at most: 2 in-edges and 2 out-edges, because in the worst case there is one in-edge imposed by program order and one in-edge imposed by log order, and likewise for out-edges. So an upper-bound on the number of edges is $2 \cdot X + 4 \cdot Y + Z$ (which is loose, as there cannot be more than $Y$ "log" edges). Summing, $O(V + E) = O(X + Y + Z)$, as claimed.

*Space complexity.* The trace *Tr* and reports are $O(X)$ and $O(Y)$, respectively; *R.M* is $O(X)$. The space of the graph $G$ is proportional to the sum of vertices and edges, which as established above is $O(X + Y + Z)$. Finally, *OpMap* is $O(Y)$ because there is one entry for each state operation. □

**Lemma 12.** *CreateTimePrecedenceGraph adds the minimum number of edges sufficient to capture the $<_{Tr}$ relation.*

*Proof.* The argument is very similar to Theorem 5 in the full version of Anderson et al. [14]; we rehearse it here.

Define the set of edges OPT as the minimum-sized set of edges in $G_{Tr}$ such that for all requests $r_1, r_2$: $r_1 <_{Tr} r_2 \iff$ there is a directed path in OPT from $r_1$ to $r_2$. We want to establish that the set of edges added by CreateTimePrecedenceGraph, call it $E$, is a subset of OPT.

If not, then there is an edge $e \in E$ but $e \notin$ OPT; label the vertices of $e$ as $r_1$ and $r_2$. Because $e \in E$, Lemma 2 implies that $r_1 <_{Tr} r_2$. But this implies, by definition of OPT, that there is a directed path from $r_1$ to $r_2$ in OPT. Yet, $e \notin$ OPT, which implies that there is at least one other request $r_3$ such that there are directed paths in OPT from $r_1$ to $r_3$ and from $r_3$ to $r_2$. This in turn means, again by definition of OPT, that

$$r_1 <_{Tr} r_3 <_{Tr} r_2.$$

However, if this is the case, then $r_3$ would have evicted $r_1$ (or an intermediate request) from the frontier by the time that $r_2$ arrived. Which implies that $r_2$ could not have been connected to $r_1$ in $E$. This is a contradiction. □

We established that $E$, which captures the relation $<_{Tr}$ (per Lemma 2), is a subset of OPT. Yet, OPT is the smallest set of edges needed to capture the relation. Therefore, $E$ and OPT are equal.